\documentclass[superscriptaddress,prd,twocolumn]{revtex4}

\usepackage{microtype}
\usepackage{amsmath}
\usepackage{amssymb}
\usepackage{amsfonts}
\usepackage{physics}
\usepackage{bm}
\usepackage[dvipsnames]{xcolor}
\usepackage{graphicx}
\newcommand{\ba}{\begin{eqnarray}}
\newcommand{\ea}{\end{eqnarray}}	

\graphicspath{{Figures/}}
\usepackage[pdftex]{hyperref}
\hypersetup{colorlinks=true, linkcolor=Blue, citecolor=purple}

\begin{document}

\title{Entanglement harvesting with coherently delocalized matter}

\author{Nadine Stritzelberger}
\thanks{nadine.stritzelberger@cantab.net}
\affiliation{Institute for Quantum Computing, University of Waterloo, Waterloo, ON N2L 3G1, Canada}
\affiliation{Department of Applied Mathematics, University of Waterloo, Waterloo, ON N2L 3G1, Canada}
\affiliation{Perimeter Institute for Theoretical Physics,
Waterloo, ON N2L 2Y5, Canada}
\affiliation{Centre for Quantum Computation and Communication Technology, School of Mathematics and Physics, The University of Queensland, St. Lucia, QLD 4072, Australia}

\author{Laura J. Henderson}
\thanks{l7henderson@uwaterloo.ca}
\affiliation{Institute for Quantum Computing, University of Waterloo, Waterloo, ON N2L 3G1, Canada}
\affiliation{Department of Physics and Astronomy, University of Waterloo, Waterloo, ON N2L 3G1, Canada}
\affiliation{Centre for Quantum Computation and Communication Technology,
School of Science, RMIT University, Melbourne, Victoria 3001, Australia}

\author{Valentina Baccetti}
\affiliation{Centre for Quantum Computation and Communication Technology,
School of Science, RMIT University, Melbourne, Victoria 3001, Australia}

\author{Nicolas C. Menicucci}
\affiliation{Centre for Quantum Computation and Communication Technology,
School of Science, RMIT University, Melbourne, Victoria 3001, Australia}

\author{Achim Kempf}
\affiliation{Institute for Quantum Computing, University of Waterloo, Waterloo, ON N2L 3G1, Canada}
\affiliation{Department of Applied Mathematics, University of Waterloo, Waterloo, ON N2L 3G1, Canada}
\affiliation{Perimeter Institute for Theoretical Physics,
Waterloo, ON N2L 2Y5, Canada}
\affiliation{Department of Physics  and Astronomy, University of Waterloo, Waterloo, ON N2L 3G1, Canada}
\affiliation{Centre for Quantum Computation and Communication Technology, School of Mathematics and Physics, The University of Queensland, St. Lucia, QLD 4072, Australia}

\begin{abstract}

We study entanglement harvesting for matter systems such as atoms, ions or molecules, whose center of mass degrees of freedom are quantum delocalized and which couple to a relativistic quantum field. 
We employ a generalized Unruh-deWitt detector model for the light-matter interaction, and we investigate how the coherent spreading of the quantum center of mass wave function of two delocalized detector systems impacts their ability to become entangled with one another, via their respective interaction with a quantum field.
For very massive detectors with initially highly localized centers of mass, we recover the results of entanglement harvesting for pointlike Unruh-deWitt detectors with classical center of mass degrees of freedom.
We find that entanglement harvesting is Gaussian suppressed in the initial center of mass delocalization of the detectors.
We further find that spatial smearing profiles, which are commonly employed to model the finite size of atoms due to their atomic orbitals, are not suited to model center of mass delocalization.
Finally, for coherently delocalized detectors, we compare entanglement harvesting in the vacuum to entanglement harvesting in media. We find that entanglement harvesting is significantly suppressed in media in which the wave propagation speed is much smaller than the vacuum speed of light.
\end{abstract}

\maketitle

\section{Introduction}

It is a well known result that first quantized, localizable matter systems can become entangled with one another via their respective interaction with the electromagnetic vacuum \cite{sorkin, Srednicki,  Steeg_Menicucci_harvesting, Edu_Alejandro_harvesting2,Edu_Alejandro_harvesting1,Edu_Achim_farming,Salton_2015, martin2013sustainable}. This process, commonly referred to as vacuum entanglement harvesting, occurs due to the fact that the vacuum state of a quantum field is entangled \cite{vacuum_entangled1, vacuum_entangled2, vacuum_entangled3, vacuum_entangled4, Nambu}. 
Entanglement harvesting has been studied extensively within the Unruh-deWitt (UdW) detector model. Despite its simplicity, the UdW detector model qualitatively captures many aspects of the light-matter interaction \cite{Unruh_UdW_detector, DeWitt_UdW_detector, birrell_davies}.
On the one hand, it models matter systems as two-level detector systems with classical center of mass degrees of freedom. On the other hand, it replaces the electromagnetic quantum field by a simpler scalar quantum field. 
Entanglement harvesting has been studied for UdW detectors not only interacting with the vacuum, but also for instance with general coherent field states \cite{All_coherent_states_entangle_equally, thermal_and_squeezed_coherent_states}, with quantum fields in curved spacetimes \cite{Steeg_Menicucci_harvesting, cosmological_quantum_entanglement,entanglement_in_curved_spacetimes,harvesting_BH,Ball_Schuller} and with quantum fields in spacetimes with non-trivial topology \cite{concurrence_idential_atoms,Ng_topology}. 

The process of entanglement harvesting depends very sensitively on the detector details \cite{Edu_Alejandro_harvesting1,Edu_Alejandro_harvesting2,Vacuum_Harvesting_experiment1,Vacuum_Harvesting_experiment2,Edu_UdW_model,No_go_entanglement_extraction}. For instance, it depends on whether or not the detectors are assumed to have a finite spatial extent, which is oftentimes incorporated into the UdW detector model by equipping the detectors with classical smearing functions \cite{Edu_UdW_model}.
Thus, for the purpose of studying entanglement harvesting in the light-matter interaction, it is crucial to let the details of the detector model reflect the physics of realistic matter systems, such as atoms, molecules or ions. The interaction of such matter systems and light is affected by two main aspects. 
First, matter systems such as atoms, molecules or ions can have a finite spatial extent due to their internal degrees of freedom, that is, due to their orbital wave functions. Second, their center of mass degrees of freedom can be quantum delocalized. While classical smearing functions have been shown to be appropriate for the purpose of modelling atomic orbitals \cite{Edu_Alejandro_harvesting2}, they are not suited to model center of mass delocalization \cite{Stritzelberger_Kempf}.

The aim of this paper is to investigate how the quantum nature of the center of mass degrees of freedom of matter affects the process of entanglement harvesting.
In the first section of this paper, we review the process of entanglement harvesting for both pointlike and spatially smeared UdW detectors, which interact with a scalar quantum field. 
In the second section, we then employ a generalized UdW detector model \cite{Stritzelberger_Kempf}, such as to describe the interaction between quantum delocalized, first quantized matter systems and a second quantized field. 
We consider two quantum delocalized detectors in their respective ground states, and we ask how their ability to become entangled with each other is affected by their mass and their initial center of mass delocalization. We recover the results of vacuum entanglement harvesting for two pointlike UdW detectors, in the limit of very large detector masses and very sharply localized center of mass degrees of freedom. 
We find that there is however no limit in which one recovers the results of vacuum entanglement harvesting for two spatially smeared UdW detectors. 
Further, we find that vacuum entanglement harvesting is suppressed in the initial delocalization, and that very delocalized detectors can not harvest any entanglement from the vacuum.

Finally, in the third section of this paper, we briefly discuss entanglement harvesting for delocalized detectors in media---which might be of interest not only for the purpose of experimentally observing entanglement harvesting, but also to potentially make use of entanglement harvesting in quantum technologies. For the sake of simplicity, we model a medium as a scalar quantum field, whose wave propagation speed differs from the vacuum propagation speed of light.
As the wave propagation speed decreases, we find that less entanglement can be harvested from the phononic ground state. 
Intuitively, this is because the phononic ground state transforms non-trivially under a quantum reference frame transformation into the rest frame of the coherently delocalized detectors, subjecting the detectors to noisy excitations which make it more difficult for the detectors to harvest entanglement.
We conjecture that matter systems are less likely to become entangled with each other by interacting with a medium than they are by interacting with the electromagnetic field.

\section{Review: Entanglement harvested by UdW detectors from the vacuum}

Before taking into account the quantum delocalization of the centers of mass of the detectors, we briefly review the process of vacuum entanglement harvesting for detectors whose center of mass degrees of freedom are described classically.
We consider two UdW detectors, labeled by $J=A,B$, which interact with a scalar quantum field.
We assume that the two detectors have the same energy gap $\Omega$ and that their classical centers of mass are located respectively at the positions $\mathbf{x}_J$. We let 
$S:=|\mathbf{x}_A - \mathbf{x}_B|$ denote the center of mass separation of the two detectors. 
We work within the interaction picture, in which operators evolve according to the free Hamiltonian 
\ba
    \hat{H}_{0} 
&=& 
    \sum_{J=A,B}
    \Omega \ket{e_J}\bra{e_J} 
    + \int d^3 k \, c k \, \hat{a}_{\mathbf{k}}^{\dagger} \hat{a}_{\mathbf{k}}^{\vphantom \dagger}
    \,. \quad 
\ea
We here let $\ket{e_J}$ and $\ket{g_J}$ denote the respective excited and ground energy eigenstates of the two detectors. The operators $\hat{a}_{\mathbf{k}}^{\vphantom \dagger}$ and $\hat{a}_{\mathbf{k}}^{\dagger}$ are the annihilation and creation operators of scalar field modes of momentum $\mathbf{k}$. We classically model the extended spatial profile of the detectors (as is common practice in the literature, see, e.g., \cite{Edu_UdW_model}) by introducing smearing profiles $\xi(\mathbf{x}- \mathbf{x}_J)$ for the two detectors.
We now assume that the monopole moment operators of the two detectors, $\hat{\mu}_J= \ket{e_J}\bra{g_J}+\ket{g_J}\bra{e_J}$, couple to the field operators $\hat{\phi}(\mathbf{x})$ via the linear monopole moment operator coupling.
In the interaction picture, the state of the system then evolves in time according to the interaction Hamiltonian
\ba
    \hat{H}_{int}(t) 
&=& 
    \sum_{J=A,B}
    \hat{H}_J(t) 
    \,,
\ea
where the interaction Hamiltonians $\hat{H}_J(t)$ respectively capture the interaction of the two detectors with the quantum field:
\ba
    \hat{H}_{J}(t)
&:=&
    \lambda \chi(t)\hat{\mu}_{J}(t) 
    \int d^3 x \,
    \xi (\mathbf{x} - \mathbf{x}_J) \hat{\phi}(\mathbf{x},t)
\ea
We here let $\lambda$ denote the interaction strength, and we introduced a switching function $\chi(t)$, via which the interaction between the detectors and the quantum field may be switched on and off.
The monopole moment operators and the field operators evolve according to their free Hamiltonians as follows:
\ba
    \hat{\mu}_J(t) 
&=& 
    e^{i\Omega t}\ket{e_J}\bra{g_J} + \text{h.c.} 
    \,,
    \\
    \hat{\phi}(\mathbf{x},t) 
&=& 
    \int \frac{d^3k}{(2\pi)^{3/2}} \sqrt{\frac{ c^2}{2 k}} 
    \Big[ e^{-ick t + i\mathbf{k}\mathbf{x}}\hat{a}_{\mathbf{k}}+ \text{h.c.} \Big] 
    \qquad 
    \label{eq:free_evolution_monopole_and_field}
\ea 

Perturbatively in the interaction strength, we can now study how much the two detectors become entangled with each other, via their respective interaction with the quantum field (see, e.g., \cite{Edu_Alejandro_harvesting2, Edu_Alejandro_harvesting1} for a derivation and in-depth discussion of what follows in this paragraph).
To this end, we consider the initial state 
$\rho(0)=
\ket{g_A}\bra{g_A}
\otimes\ket{g_B}\bra{g_B} 
\otimes \ket{0}\bra{0}$, for which the two detectors are unentangled and the quantum field is in its vacuum state $\ket{0}$. 
Evolving the initial state in time and taking the partial trace over the field degrees of freedom, one obtains the partial state $\rho_{AB}(t)$ of the two detectors after their interaction with the quantum field. Employing the basis 
$\{ \ket{g_A}\ket{g_B}, \ket{g_A}\ket{e_B}, \ket{e_A}\ket{g_B}, \ket{e_A}\ket{e_B} \}$, the partial state of the detectors becomes
\ba
\rho_{AB}(t)
&=:& 
\begin{pmatrix}
1- 2P^{\,c} & 0 & 0 & \mathcal{M}^{\,c*} \\
0 & P^{\,c} & \mathcal{L}^{\,c} & 0 \\
0 & \mathcal{L}^{\,c} & P^{\,c} & 0 \\
\mathcal{M}^{\,c} & 0 & 0 & 0
\end{pmatrix}
+ \mathcal{O}(\lambda^4) \,,
\qquad 
\ea
to second perturbative order in the interaction strength. 
Here, $P^{\,c}$ denotes the excitation probability of the two detectors respectively, and $\mathcal{M}^{\,c}$ has traditionally been referred to as the entangling term. The excitation probabilities of the two detectors are equal, since we here consider the same switching function and the same smearing profile (up to a displacement in space) for the two detectors.
We here let the superscript $c$ remind us of the classical nature of the center of mass degrees of freedom of the two UdW detectors.

To measure the entanglement between the internal degrees of freedom of the two detectors, we will here employ the entanglement negativity \cite{negativity}, which is an entanglement monotone \cite{negativity_entanglement_monotone, Lee_2000}. The entanglement negativity is defined for a density matrix as the absolute value of the sum of the negative eigenvalues of the partially transposed density matrix. To second perturbative order in the interaction strength, the negativity for the partial state of the two detectors becomes (see, e.g., \cite{Edu_Alejandro_harvesting2, Edu_Alejandro_harvesting1})
\ba
    \mathcal{N}^{\,c}
&=& 
    \max \big\{ 0 \,,\,  - P^{\,c} + |\mathcal{M}^{\,c}| \big\} 
    \,.
\ea

It is worth mentioning here that a rich variety of quantitative entanglement measures have been established in the literature (see, e.g., \cite{entanglement_measures_review}). 
For instance, we could just as well use concurrence \cite{concurrence} as our measure for entanglement. For two identical detectors and to second perturbative order, concurrence and entanglement negativity have however been shown to be equivalent entanglement measures \cite{concurrence_idential_atoms}. For the purpose of this paper, we will restrict our attention to entanglement negativity. 

Let us now consider a switching function of compact support of the form
\ba
    \chi(t) 
&=&
    \begin{cases}
    \sin\left(t / \sigma \right) ,& \text{for} \quad 0 \leq t \leq \pi\sigma 
    \\
    0  & \text{otherwise} \,.
    \end{cases}
    \label{eq:sine_switching}
\ea
Examples for the use of compact switching functions can be found e.g. in \cite{Henderson:2020zax, Edu_Alejandro_harvesting1}. We employ a compact switching function in order to ensure that the interaction between the detectors and the field is switched on only during a compact time interval, $t\in [0,\pi\sigma]$. The importance of employing a compact switching function will become apparent in the next section. Integrating over all times, the excitation probabilities and the entangling term become
\ba
    P^{\,c}
&=&
    \frac{\lambda^2 \sigma^2}{(2\pi)^3}
    \int \frac{d^3 k}{2ck} \,
    |\widetilde{\xi}(\mathbf{k})|^2 \, A(k)
    \,, 
    \\
    \mathcal{M}^{\,c}
&=&
    - \frac{\lambda^2\sigma^2 e^{i\pi\Omega\sigma} }{c}
    \int \frac{d^3 k}{k}
    \frac{e^{2i\mathbf{k}\cdot \mathbf{x}_0} \, \widetilde{\xi}(\mathbf{k})^2 B(k) }
    {1-\sigma^2 (\Omega+ck)^2} 
    \,, \qquad
\ea
where $\widetilde{\xi}(\mathbf{k})$ denotes the Fourier transformation of the spatial smearing profile $\xi(\mathbf{x})$ and where we defined the following functions:
\ba
   A(k)
&:=&
    \frac{1 + \cos(\pi\sigma(\Omega + ck))}
    { \left(\sigma^2 (\Omega + ck)^2  - 1 \right)^2 }
    \,,
    \\
    B(k)
&:=&
    \frac{i(2\Omega+ck) \sin(\pi\Omega\sigma)}{2\Omega (1-\Omega^2\sigma^2)} 
    + \frac{e^{-i\pi\Omega\sigma}+e^{-i\pi ck\sigma}}{1-\sigma^2(\Omega-ck)^2}
    \qquad \,\,
\ea
\begin{figure}
    \includegraphics[width=0.44\textwidth]{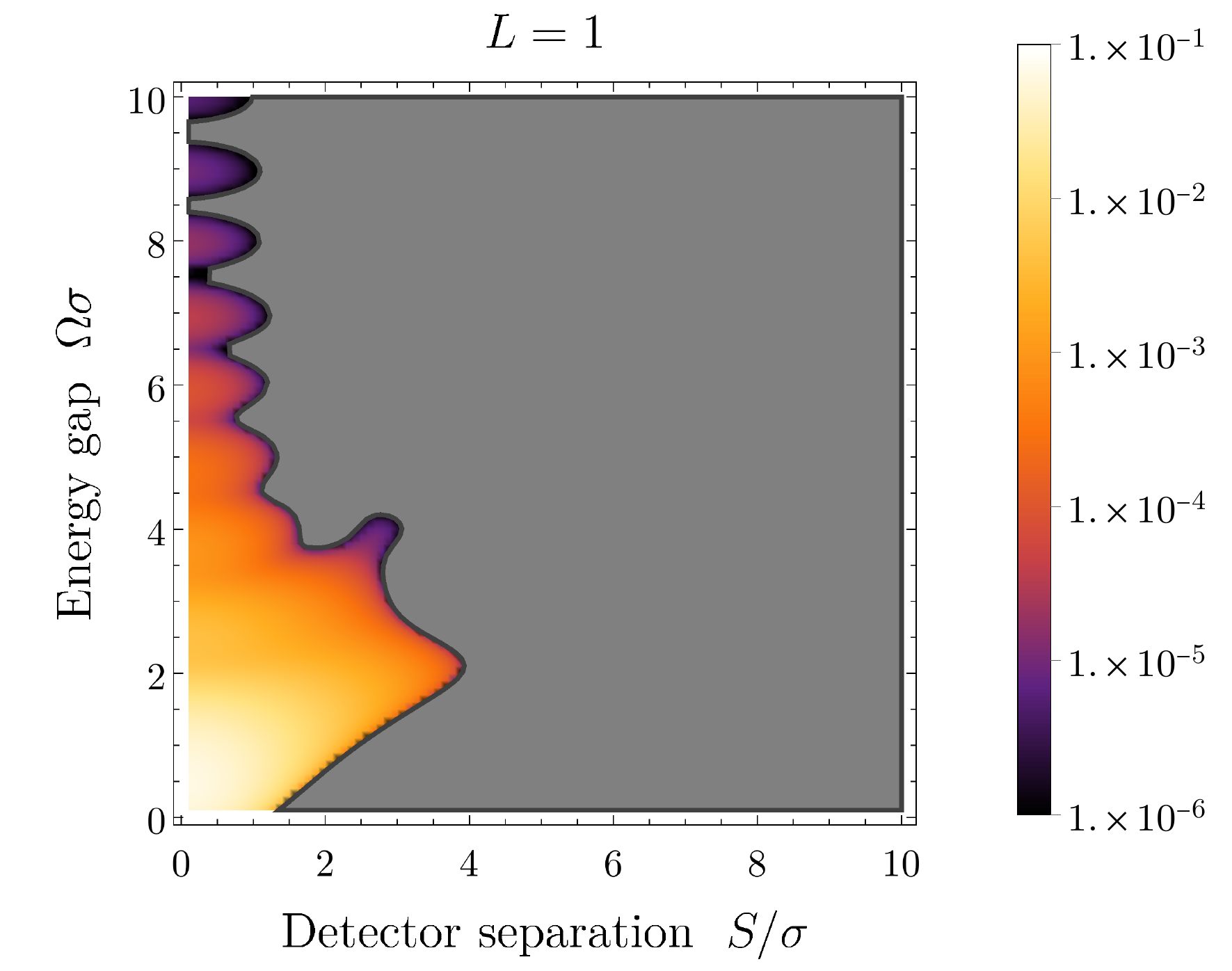}
    \par \vspace{0.8cm}
    \includegraphics[width=0.44\textwidth]{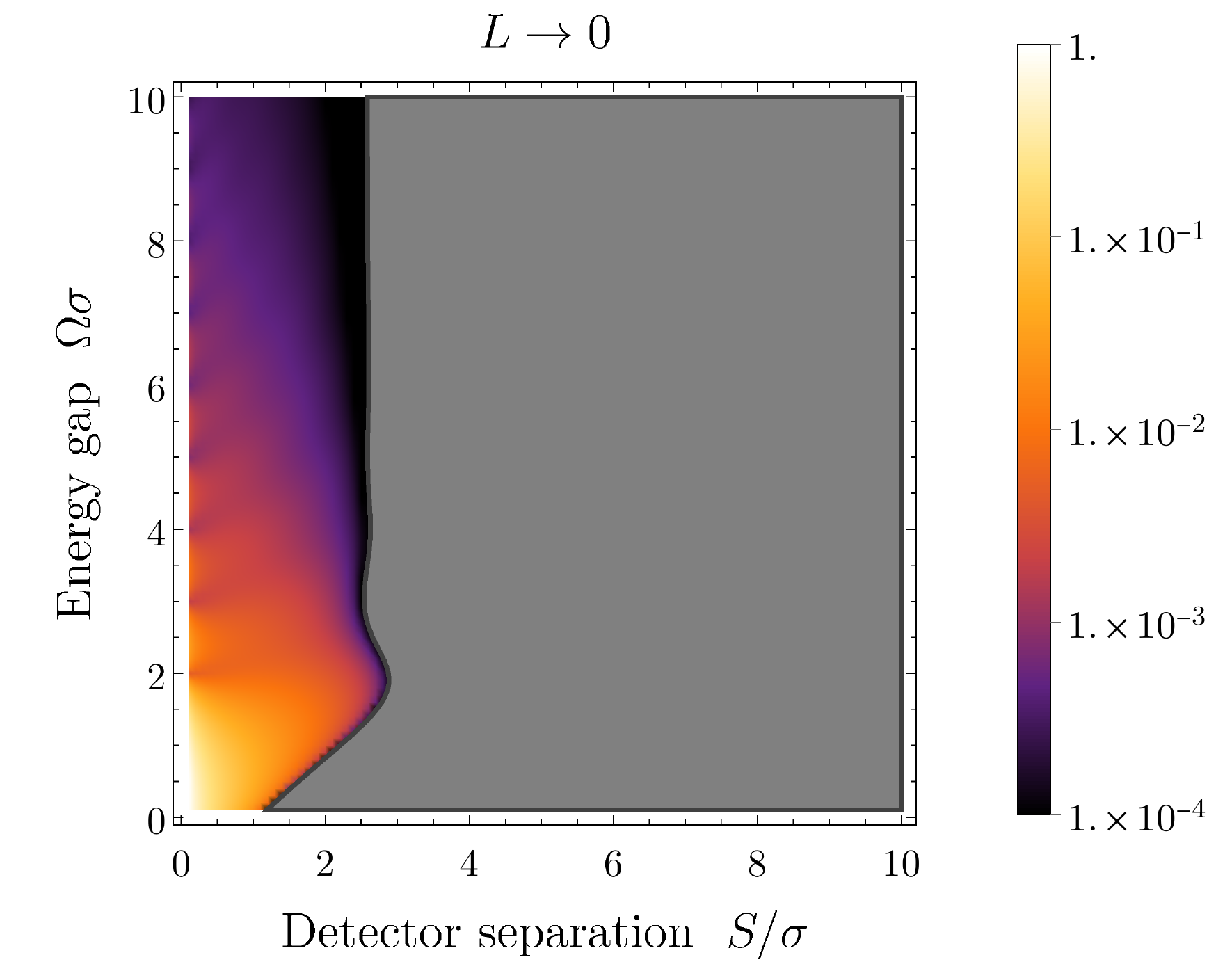}
    \par \vspace{0.2cm}
    \caption{%
        The negativity $\mathcal{N}$, as a function of the energy gap $\Omega$ and the separation $S$ of the detectors, plotted \emph{(top)} 
        for spatially smeared UdW detectors
        with $L=\sigma$ and
        \emph{(bottom)} for pointlike UdW detectors. The regions of zero negativity are  marked in grey.
        }
    \label{fig:Negativity_UdW_detectors}
\end{figure}
We can now specify a spatial smearing function according to which the detectors couple to the field. We could for instance model the spatial extent of the detectors via Gaussian smearing profiles of standard deviation $L/2$:
\ba
    \xi(\mathbf{x}) 
&=&
    \left( \frac{2}{\pi L^2} \right)^{3/2} 
    e^{- 2 \mathbf{x}^2/L^2}
    \label{eq:spatial_smearing}
\ea
We will here refer to $L$ as the ``width" of the smearing profiles.
Since the smearing profiles are normalized, we may interpret them as classical probability distributions, according to which the detectors couple to the quantum field.
We find that the excitation probabilities and the entangling term depend on the width $L$ of the smearing profiles, the energy gap $\Omega$ of the detectors and the total interaction time $\pi\sigma$, and the entangling term additionally depends on the separation $S$ of the two detectors:
\ba
    P^{\,c}_{L}
&=&
    \frac{\lambda^2 \sigma^2}{2\pi^2 c}
    \int_0^{\infty} dk \, k \,
    e^{-\frac{L^2 k^2}{4}}
    A(k) 
    \,, \label{eq:PA_classical_Gaussian_smeared}\\
    \mathcal{M}^{\,c}_{L}
&=& 
    - \frac{\lambda^2\sigma^2 e^{i\pi\Omega\sigma} }{2\pi^2 cS}
    \int_{0}^{\infty} dk \, 
    \frac{\sin(kS) e^{-\frac{L^2 k^2}{4}} B(k) }{1-\sigma^2 (\Omega+ck)^2}
    \label{eq:M_classical_Gaussian_smeared}
    \qquad
\ea
We here let the subscript $L$ indicate that we employed Gaussian smearing profiles.
In the limit of very sharply peaked smearing profiles ($L\to 0$), we recover the excitation probabilities and the entangling term for pointlike UdW detectors:
\ba
    P^{\,c}_{0}
&=& 
    \frac{\lambda^2 \sigma^2}{2\pi^2 c}
    \int_0^{\infty} dk \, k \,
    A(k)
    \label{eq:PA_classical_pointlike}\,,
    \\
    \mathcal{M}^{\,c}_{0}
&=& 
    - \frac{\lambda^2\sigma^2 e^{i\pi\Omega\sigma} }{2\pi^2 cS}
    \int_{0}^{\infty} dk \, 
    \frac{\sin(kS) B(k) }{1-\sigma^2 (\Omega+ck)^2}
    \label{eq:M_classical_pointlike}
    \qquad 
\ea

The negativity for both pointlike and Gaussian smeared UdW detectors is plotted in Fig.(\ref{fig:Negativity_UdW_detectors}), as a function of the energy gap $\Omega$ of the detectors and their separation $S$.
We can see that the negativity decreases with increasing widths $L$ of the Gaussian smearing profiles. Intuitively, we can understand this behaviour as follows: the field amplitudes at different points in space and time are quantum correlated, which is why two spatially separated detectors can become entangled with one another in the first place. 
Spatially smeared UdW detectors average the quantum field fluctuations over extended spatial regions, and the larger these spatial regions are, the less entanglement the detectors can harvest from the quantum field.
In the negativity plots in Fig.(\ref{fig:Negativity_UdW_detectors}) we further observe a resonance-like behaviour, for energy gaps which are multiples of the switching scale $\sigma$, which manifests itself in slight ripples in the negativity for pointlike UdW detectors, and more pronounced oscillations for spatially smeared UdW detectors.

\section{Entanglement harvested by coherently delocalized detectors from the vacuum}

The results for entanglement harvesting in the previous section relied on the assumption that the center of mass degrees of freedom of the matter systems under investigation are classical. In this section, we want to study how the process of entanglement harvesting is affected when the center of mass motion of the two detectors is quantum uncertain. 

A possible setup we have in mind here is the following:
Let us consider two atoms, which are initially localized in a certain region of space via a center of mass position measurement, and which are then left to evolve freely. The center of mass wave functions of the two initially localized atoms spread and the atoms dynamically delocalize in space. 
Let us now imagine that these two coherently delocalizing atoms interact with the electromagnetic vacuum. How much will the internal degrees of freedom of the atoms become entangled with each other? How will the result depend on the mass of the detectors, their initial localization, and their dynamical delocalization process? How will the results compare to the results for entanglement harvesting with classical center of mass degrees of freedom?

To study these questions, we employ a generalized UdW detector model, as described in \cite{Stritzelberger_Kempf}: We again replace the electromagnetic field by a simpler scalar field, and we model the atoms as two-level detector systems. 
However, in order to allow the detectors to coherently delocalize, we now consider the center of mass degrees of freedom of the two detectors to be quantized. We let $\hat{\mathbf{x}}_J$ and $\hat{\mathbf{p}}_J$ denote the center of mass position and momentum operators of the two detectors. We will here assume that the two detectors are of equal mass $M$. 
In the interaction picture, operators then evolve according to the free Hamiltonian
\ba
    \hat{H}_{0} 
&=& 
    \sum_{J=A,B}
    \left[
    \frac{\hat{\mathbf{p}}_J^{\,\,2}}{2M} 
    + \Omega \ket{e_J}\bra{e_J} 
    \right]
    + \int d^3 k \, c k \, \hat{a}_{\mathbf{k}}^{\dagger} \hat{a}_{\mathbf{k}}^{\vphantom \dagger}
    \,.
    \qquad 
\ea
We again couple the detectors to the quantum field via the monopole moment coupling. However, the field operators now take the center of mass position operators of the two detectors as their argument:
\ba
    \hat{H}_{int}(t) 
&=& 
    \lambda \chi(t) \sum_{J=A,B}
    \hat{\mu}_J(t)
    \hat{\phi}(\hat{\mathbf{x}}_J,t)
\ea
We note that the interaction Hamiltonian involves functions $\hat{\phi}(\hat{\mathbf{x}}_J,t)$ which are dependent on operators and which are operator valued themselves.
An interaction of this form can be made sense of via the spectral theorem of functional calculus \cite{functional_calculus, maria_thesis, Stritzelberger_Kempf} as follows,
\ba
    \hat{\phi}(\hat{\mathbf{x}}_J,t)
&:=&
    \int d^3 x_J \,
    \hat{\mathcal{P}}(\mathbf{x}_J,t)
    \hat{\phi}(\mathbf{x}_J,t) \,,
    \qquad 
\ea
where the operators $\hat{\mathcal{P}}(\mathbf{x}_J,t)$ are given by
\ba
    \hat{\mathcal{P}}(\mathbf{x}_J,t)
&:=&
    \ket{\mathbf{x}_J(t)}\bra{\mathbf{x}_J(t)}
    \\
&=&
    \int \frac{d^3 p \, d^3 q}{(2\pi)^3}
    e^{-i(\mathbf{p} - \mathbf{q} )\cdot \mathbf{x}_J + it\frac{ p^2 - q^2 }{2M}} 
    \ket{\mathbf{p}} \bra{\mathbf{q}}
    \,.
    \qquad 
\ea 
We again assume that initially the two detectors are in their ground states and the field is in its vacuum state. We further let $\ket{\varphi_J}$ denote the initial center of mass states of the two detectors. 
The initial state of the system then reads
$\rho(0) = 
\ket{\varphi_A} \bra{\varphi_A}
\otimes \ket{g_A}\bra{g_A}
\otimes 
\ket{\varphi_B} \bra{\varphi_B}
\otimes \ket{g_B}\bra{g_B}
\otimes \ket{0}\bra{0}$.
We can express the initial center of mass states both in terms of the initial center of mass wave functions in the position and in the momentum representation:
\ba
    \ket{\varphi_J} 
&=& 
    \int d^3 x \, 
    \varphi_J(\mathbf{x}) 
    \ket{\mathbf{x}}
    = 
    \int d^3 p \, 
    \widetilde{\varphi}_J(\mathbf{p}) 
    \ket{\mathbf{p}} 
    \qquad 
\ea
We again want to investigate how entangled the internal degrees of freedom become with each other via the interaction of the detectors with the quantum field. To this end, we evolve the initial state in time and trace over both the field and the center of mass degrees of freedom. To second perturbative order, we again obtain the partial state of the two detectors:
\ba
\rho_{AB}(t)
&=:& 
\begin{pmatrix}
1- P_A - P_B & 0 & 0 & \mathcal{M}^* \\
0 & P_B & \mathcal{L} & 0 \\
0 & \mathcal{L} & P_A & 0 \\
\mathcal{M} & 0 & 0 & 0
\end{pmatrix}
+ \mathcal{O}(\lambda^4)
\qquad 
\ea
We again calculate the entanglement negativity for this density matrix, as a quantitative measure for how much entanglement the internal degrees of freedom of the two coherently delocalizing detectors can harvest from the vacuum:
\ba
    \mathcal{N} 
&=& 
    \max \Bigg\{ 0 \,,\, 
    \sqrt{ \tfrac{(P_A-P_B)^2}{4}
    + |\mathcal{M}|^2 }
    \, - \tfrac{P_A + P_B}{2}
    \Bigg\} \qquad 
\ea
We find the excitation probabilities $P_J$ of the two detectors to be 
\ba
    P_J
&:=&
    \frac{\lambda^2}{(2\pi)^3} 
    \int_{0}^{t} dt_1 
    \int_{0}^{t} dt_2 \,
    \chi(t_1)\chi(t_2)
    \int d^3p 
    \nonumber
    \\
    && \times 
    \int \frac{d^3 k}{2ck}
    \left| \widetilde{\varphi}_J(\mathbf{p}) \right|^2 
    e^{i(t_1-t_2) \left( \Omega + ck + \frac{ k^2 - 2\mathbf{p} \cdot \mathbf{k} }{2M} \right)}
    \,,\qquad 
    \label{eq:PA_general}
\ea
and we note that the excitation probabilities of the two detectors depend on their center of mass states only via their respective momentum probability distributions $|\widetilde{\varphi}_J(\mathbf{p})|^2$. Unlike the excitation probabilities, we find that the entangling term $\mathcal{M}$ also depends on the phases of the initial center of mass momentum wave functions:
\ba
    \mathcal{M} 
&&:=
    - \lambda^2
    \int_{0}^{t} dt_1 
    \int_{0}^{t_1} dt_2 \,  
    \chi(t_1)\chi(t_2) 
    \int \frac{d^3 p_1 d^3 p_2}{(2\pi)^3} 
     \nonumber
    \\
    && \times
    \int \frac{d^3 k}{2ck} 
    \, \widetilde{\varphi}_A(\mathbf{p_1}+ \mathbf{k}) 
    \, \widetilde{\varphi}_B(\mathbf{p_2}-\mathbf{k}) 
    \, \widetilde{\varphi}_{A}^{\,*}(\mathbf{p_1})
    \, \widetilde{\varphi}_{B}^{\,*}(\mathbf{p_2})
    \nonumber
    \\
    && \times 
    e^{i(t_1+t_2)\left( \Omega - \frac{\mathbf{k}(\mathbf{p_1}-\mathbf{p_2})}{2M} - \frac{k^2}{2M}\right)}
    \bigg[
        e^{ -i (t_1-t_2) \left( ck - \frac{\mathbf{k} (\mathbf{p}_1 + \mathbf{p}_2 )}{2M} \right)}
    \nonumber
    \\
    && + e^{ -i (t_1-t_2) \left( ck + \frac{\mathbf{k} (\mathbf{p}_1 + \mathbf{p}_2 )}{2M} \right)} 
    \bigg]
\ea 
The phases of the momentum wave functions carry the position information of the two wave functions, and as expected, the entangling term thus depends on the spatial locations of the two center of mass wave packets.
While the excitation probabilities of the detectors respectively only depend on the properties of one detector, the entangling term thus depends on the properties of both detectors. As suggested e.g. in \cite{Edu_Alejandro_harvesting2}, we can thus think of the excitation of the respective detectors according to $P_J$ as local noise, while the nonlocal entangling term $\mathcal{M}$ describes entangling excitations which are shared by the two detectors. 

We can now specify the initial center of mass wave functions for the two detectors. Let us here consider detectors whose center of mass position wave functions are Gaussian wave packets of initial width $L$, respectively centered around $\mathbf{x}_J$, at a separation $S:=|\mathbf{x}_A - \mathbf{x}_B|$:
\ba
    \varphi_{J}(\mathbf{x}\,) 
&=&
    \left( \frac{2}{\pi L^2} \right)^{3/4} e^{-\frac{| \mathbf{x} + \mathbf{x}_J |^2}{L^2} } 
    \,,
    \\
    \widetilde{\varphi}_{J}(\mathbf{p}\,) 
&=&
    \left( \frac{L^2}{2\pi} \right)^{3/4} e^{-\frac{p^2 L^2}{4} + i\mathbf{p} \cdot \mathbf{x}_J} 
    \label{eq:Gaussian_wavepacket_p}
\ea
The momentum probability distributions resulting from these momentum wave functions are the same for both detectors,
$|\widetilde{\varphi}_{J}(\mathbf{p}\,)|^2
= L^3/(2\pi)^{3/2} e^{-p^2 L^2/2}$. We thus find that the excitation probabilities of the two detectors are equal, $P_A=P_B=:P$, and the negativity reduces to
\ba
    \mathcal{N} 
&=& 
    \max \big\{ 0 \,,\, 
    |\mathcal{M}|
    \, - P
    \big\} 
    \,.
\ea
In order for the two detectors to harvest entanglement from the vacuum, the nonlocal entangling excitations thus need to dominate over the local excitation of the respective detectors \cite{Edu_Alejandro_harvesting2}. 

We can now see why it is important to employ a switching function of compact support. Under the free quantum mechanical time evolution, the wave packets in Eq.(\ref{eq:Gaussian_wavepacket_p}) start out completely delocalized in space for $t\to -\infty$, then flow together to Gaussians of width $L$ at time $t=0$, and then spread again into completely delocalized states for $t\to\infty$. If we employed a switching function of non-compact support, such as a Gaussian switching function, we would need to consider the completely delocalized center of mass wave packets at time $t\to -\infty$ as the initial center of mass states. However, we want to consider the localized wave packets in Eq.(\ref{eq:Gaussian_wavepacket_p}) as the center of mass states at the initial time $t=0$, since our aim here is to study how entanglement harvesting is affected by the center of mass spreading of initially localized detectors. We therefore need to ensure that the interaction with the quantum field is switched on precisely at time $t=0$, which in turn is why we need to employ a switching function of compact support.
Let us here again employ the compact sine switching function in Eq.(\ref{eq:sine_switching}). We obtain
\ba
    P
&=& 
    \frac{\lambda^2 \sigma^2}{\pi c} 
    \int_0^{\infty} dp \, p^2 \,
    |\widetilde{\varphi}_{J}(\mathbf{p}\,)|^2 \,
    \mathcal{U} (p) 
    \label{eq:PA_with_sine_switching}
\ea
for the excitation probabilities. 
We here defined the template function
\ba
    \mathcal{U}(p)
&:=&
    \int_{-1}^{1} dz \, 
    \int_0^{\infty} dk \, k \,
    \left[ 1 + \cos \left( \pi\sigma \left( F - \frac{kpz}{M} \right) \right) \right]
    \nonumber
    \\
    && \times 
    \left[\sigma^2 \left( F - \frac{kpz}{M} \right)^2  - 1 \right]^{-2}
    \,, 
    \label{eq:U}
\ea
where $p$ is the detector's recoil momentum and where we defined $F:= \Omega + ck + k^2/(2M)$ for convenience of notation. We refer to $\mathcal{U}(p)$ as a ``template function" due to the fact that the function is independent of the center of mass states of the detectors.
For the entangling term we obtain
\ba
    \mathcal{M} 
&=& 
    \frac{\lambda^2 \sigma^2 L^2}{c(2\pi)^3}
    \int_{-\infty}^{\infty} d p_1
    \int_{-\infty}^{\infty} dp_2
    \int_0^{\infty} dk  
    \frac{\sin(kS)}{S} 
    \nonumber
    \\
    && \times 
    e^{-L^2 ( p_1^2 + p_2^2)/2} 
    e^{-L^2k^2/4} \,
    \mathcal{V} (k,p_1,p_2) 
    \,. \qquad \quad 
    \label{eq:M_with_sine_switching}
\ea
We here again defined a template function,
\ba
    \mathcal{V}(k,p_1,p_2) 
:=
    \sum_{j=0}^{1}
    \frac{
    e^{i \pi  \alpha  \sigma }}{1-\sigma ^2 (\alpha +\beta_j )^2}
    \qquad \qquad\quad\quad\qquad 
    \nonumber
    \\
    \qquad
    \times 
    \Bigg(
    \frac{i (2 \alpha +\beta_j ) \sin (\pi  \alpha  \sigma )}{2 \alpha  \left(1-\alpha ^2 \sigma ^2\right)}
    +
    \frac{e^{-i \pi  \alpha  \sigma }+e^{-i \pi  \beta_j  \sigma }}{1-\sigma ^2 (\alpha -\beta_j )^2} 
    \Bigg)
    \,,\quad
    \label{eq:V}
\ea
and we defined $\alpha := \Omega - k (p_1 - p_2) /(2M)$ as well as $\beta_j := ck +(-1)^j k (p_1 + p_2) /(2M)$ for convenience of notation.

Since we work within a framework in which the center of mass dynamics are described by the Schr\"{o}dinger equation, we need to ensure that the virtual center of mass velocities are well within the non-relativistic regime. That is, we need to ensure that the momentum probability distributions $|\widetilde{\varphi}_J(\mathbf{p}\,)|^2$ have contributions only for momenta corresponding to virtual velocities much smaller than the speed of light. Let us here restrict the virtual velocities to velocities no larger than one percent of the vacuum speed of light, $v:=p/M\leq 0.01c$. The Gaussian momentum probability distributions of the detectors have a standard deviation of $1/L$. We can thus assume to a very good approximation (within $3.5$ standard deviations away from the mean) that the center of mass momenta $\mathbf{p}$ in the probability distributions satisfy $pL \lesssim 3.5$. The non-relativistic regime therefore corresponds to parameters $L$ and $M$ satisfying
\ba
    LMc \gtrsim 3.5 \times 10^2 \,.
    \label{eq:non-relativistic-regime}
\ea
The center of mass wave function of a coherently delocalized detector spreads faster for smaller $L$ and $M$, that is, for initially more sharply localized detectors and for smaller detector masses. Consequently, for a given detector mass, there is a minimal initial delocalization width we can consider while staying within the non-relativistic regime for the virtual center of mass velocities.

Provided that we chose appropriate parameters $M$ and $L$, we can now expand the template functions $\mathcal{U}$ and $\mathcal{V}$ for non-relativistic virtual center of mass velocities.
We Taylor expand $\mathcal{U}$ around $p/(Mc)=0$, and we Taylor expand $\mathcal{V}$ around both $p_1/(Mc)=0$ and $p_2/(Mc)=0$. To second order, we then obtain the following simplified expressions for the excitation probabilities and the entangling term:
\ba
    P
&=&
    \frac{\lambda^2 \sigma^2}{4 \pi^2 c} 
    \bigg[ \mathcal{U}(0) + \frac{3}{2 L^2} \frac{\partial^2 \mathcal{U}}{\partial p^2} (0) 
    + \mathcal{O}\left( (LMc)^{-4} \right) \bigg] 
    \,,\,\,\,\,
    \quad
    \label{eq:Taylor_expansion_P}
    \\
    \mathcal{M}
&=&
    \frac{\lambda^2 \sigma^2}{4\pi^2 c} 
    \int_0^{\infty} dk  
    \frac{\sin(kS)}{S} 
    e^{- L^2 k^2/4}
    \bigg[
    \mathcal{V}(k,0,0)
    \nonumber
    \\
    && 
    + \frac{1}{2L^2}\left[ \frac{\partial^2 \mathcal{V} }{\partial p_1^2} (k,0,0)
    + \frac{\partial^2 \mathcal{V} }{\partial p_2^2}(k,0,0) \right]
    \nonumber
    \\
    &&
    + \mathcal{O}\left( (LMc)^{-4} \right)
    \bigg]
    \label{eq:Taylor_expansion_M}
\ea

\begin{figure}
    \includegraphics[width=0.45\textwidth]{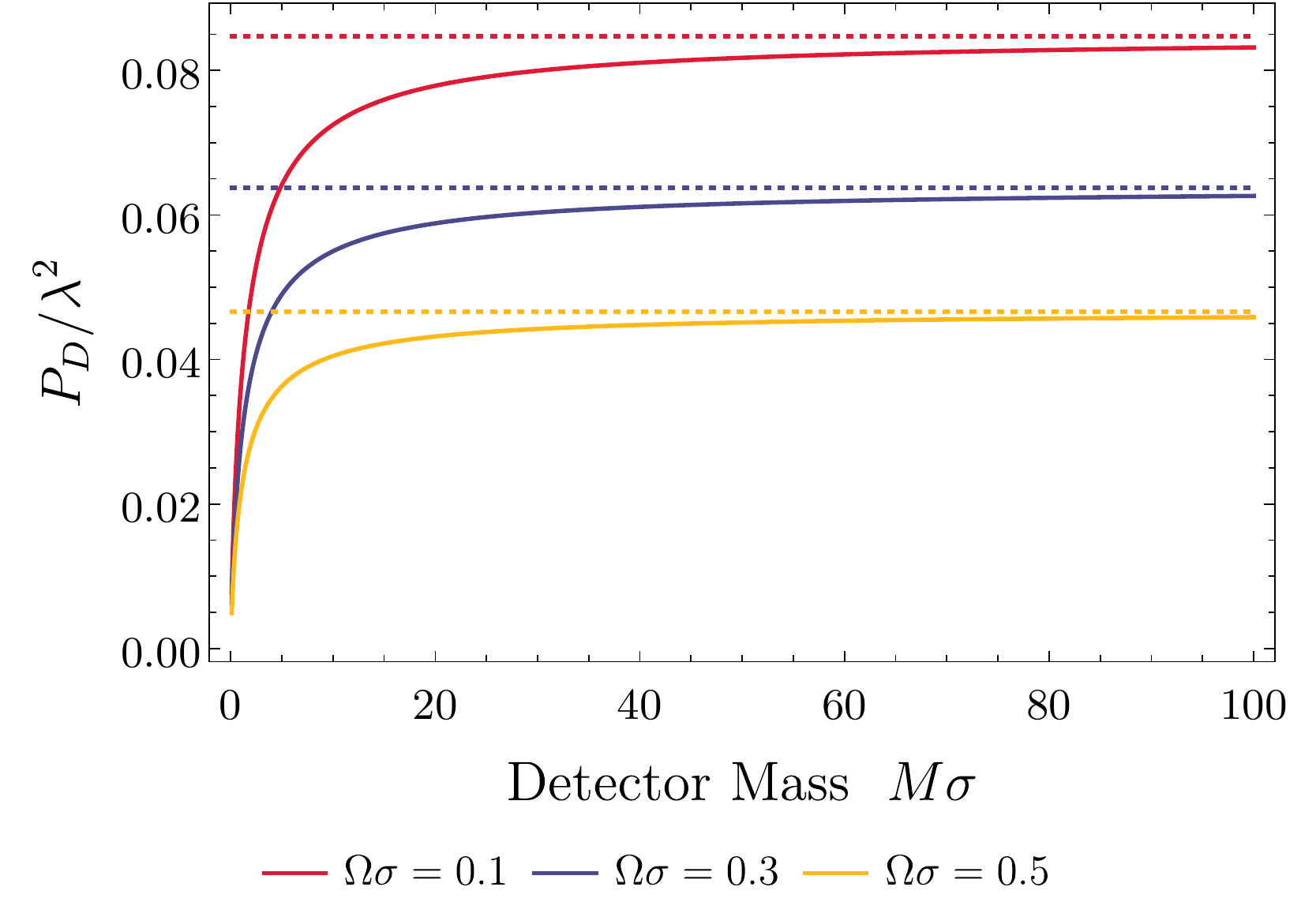} 
    \caption{The transition probability for a detector as a function of its mass, for different energy gaps and with $L=1000\sigma$.  The dotted lines represent the excitation probabilities of pointlike UdW detectors with the same energy gaps as the respective massive detectors.}
    \label{fig:P}
\end{figure}

\begin{figure}
    \par \vspace{0.2cm}
    \includegraphics[width=0.45\textwidth]{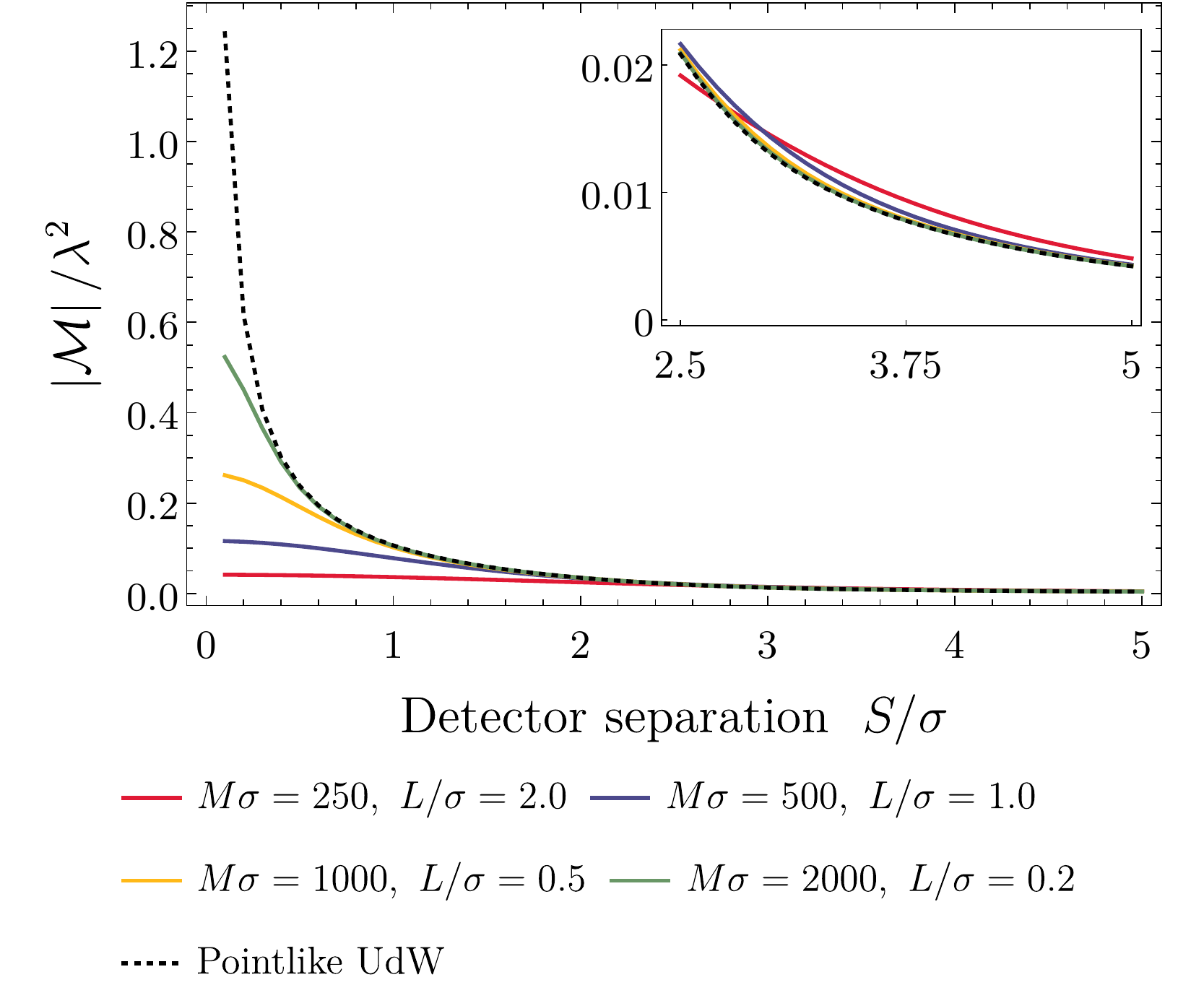} 
    \par \vspace{0.15cm}
    \caption{The entangling term $\mathcal{M}$ for pointlike UdW detectors as well as for massive detectors, as a function of the detector's separation $S$ (with $\Omega\sigma =0.1$). For the massive detectors, we chose different values for $M$ and $L$ such as to keep their product constant ($ML=500$), which fixes the virtual velocities at which the detectors dynamically delocalize. 
    }
    \label{fig:M}
\end{figure}

As displayed in Fig.(\ref{fig:P}), we find that the excitation probabilities decrease, both for increasing energy gaps $\Omega$ and decreasing detector masses $M$. 
Intuitively, this behavior can be explained as follows: 
switching the interaction on and off breaks time translation invariance and therefore provides energy for the excitation and the recoil of the detectors and the excitation of the field.
The kinetic energy of the recoil becomes larger for smaller detector masses. 
Since the excitation of the detector is always accompanied by the emission of a field quantum and the recoil of the detector, the excitation process becomes energetically more expensive for larger energy gaps and smaller detector masses.
In the limit of infinitely small energy gaps, the detectors essentially turn into simple charges and all the switching energy can go into the recoil of the detectors and the excitation of the field. 
Similarly, in the limit of infinitely large detector masses, the kinetic recoil energy tends to zero and all the switching energy can go into the excitation of the field and the internal degrees of freedom. 
For the excitation probability of a very massive detector, it is therefore justifiable to neglect the recoil of the detector and to model the center of mass degrees of freedom classically.
As can be seen in Fig.(\ref{fig:P}), the excitation probability in the limit of large detector masses indeed approaches the excitation probability of a pointlike UdW detector. 

As displayed in Fig.(\ref{fig:M}), we further find that the entangling term is suppressed both in the separation of the two detectors and in the initial center of mass delocalization widths.
Intuitively, this is because the amplitude of the quantum field fluctuations (which correlate the quantum field amplitudes at different points in space and time) decrease with the fluctuation size.
The more delocalized the detectors are initially, the larger are the spatial regions in which the two detectors 
probe the quantum field fluctuations, and the smaller is thus the entangling term. 
In fact, as we can see from Eq.(\ref{eq:Taylor_expansion_M}), the entangling term for coherently delocalized detectors is Gaussian suppressed in the initial delocalization width---contrary to the excitation probabilities, whose leading order term does not depend on the initial delocalization width at all, as we can see from Eq.(\ref{eq:Taylor_expansion_P}). Therefore, the ability of quantum delocalized detectors to harvest entanglement from the vacuum is Gaussian suppressed in the initial center of mass delocalization.  

Let us now see whether we can find a limit in which we recover the entanglement harvesting results for UdW detectors, that is, for detectors with classical center of mass degrees of freedom.
We start by exploring the negativity in the limit of very large detector masses ($M\rightarrow \infty$), while keeping the initial delocalization width $L$ fixed. 
One might expect to recover the classical behaviour of UdW detectors in this limit, since
the dynamical quantum center of mass delocalization process becomes very slow:
the virtual center of mass velocities satisfy $v\lesssim 3.5/(LM)$ and thus tend to zero in this limit. Even though the detectors each have a finite initial delocalization width, their center of mass wave packets do not coherently spread any further.
We indeed find that the excitation probabilities and the entangling term respectively reduce to the excitation probabilities and the entangling term for UdW detectors.
However, there is a twist: 
The excitation probabilities reduce to the excitation probabilities for pointlike UdW detectors, $P \to P^{\,c}_{0}$, while the entangling term reduces to the entangling term for Gaussian smeared UdW detectors,
$\mathcal{M} \to \mathcal{M}^{\,c}_{L}$. 
Thus, in the limit of very large detector masses and for finite initial delocalization widths $L$, the negativity neither reduces to the negativity for a pair of pointlike UdW detectors, nor to the negativity for a pair of Gaussian smeared UdW detectors.

Intuitively, we can understand this behavior as follows. In the infinite mass limit, the kinetic energy of the recoil of the detectors tends to zero. The center of mass degrees of freedom no longer play a role in the energy balance of the excitation process of the detectors, and the recoil of the detector becomes negligible. 
We can thus effectively interpret the center of mass probability distributions as classical probability distributions, of finite and constant width, for the positions of two pointlike UdW detectors. 
Since the excitation probability of a pointlike UdW detector is independent of the position of the detector, we recover the results for pointlike UdW detectors for the excitation probabilities. On the other hand, the nonlocal entangling excitations shared by two pointlike UdW detectors depend on the detector separation, and therefore they also depend on the classical position probability distributions for the two detectors. Consequently, the entangling term for incoherently delocalized detectors does not reduce to the entangling term for pointlike UdW detectors, but rather to the entangling term for spatially smeared UdW detectors. 

\begin{figure*}
    \includegraphics[width=0.44\textwidth]{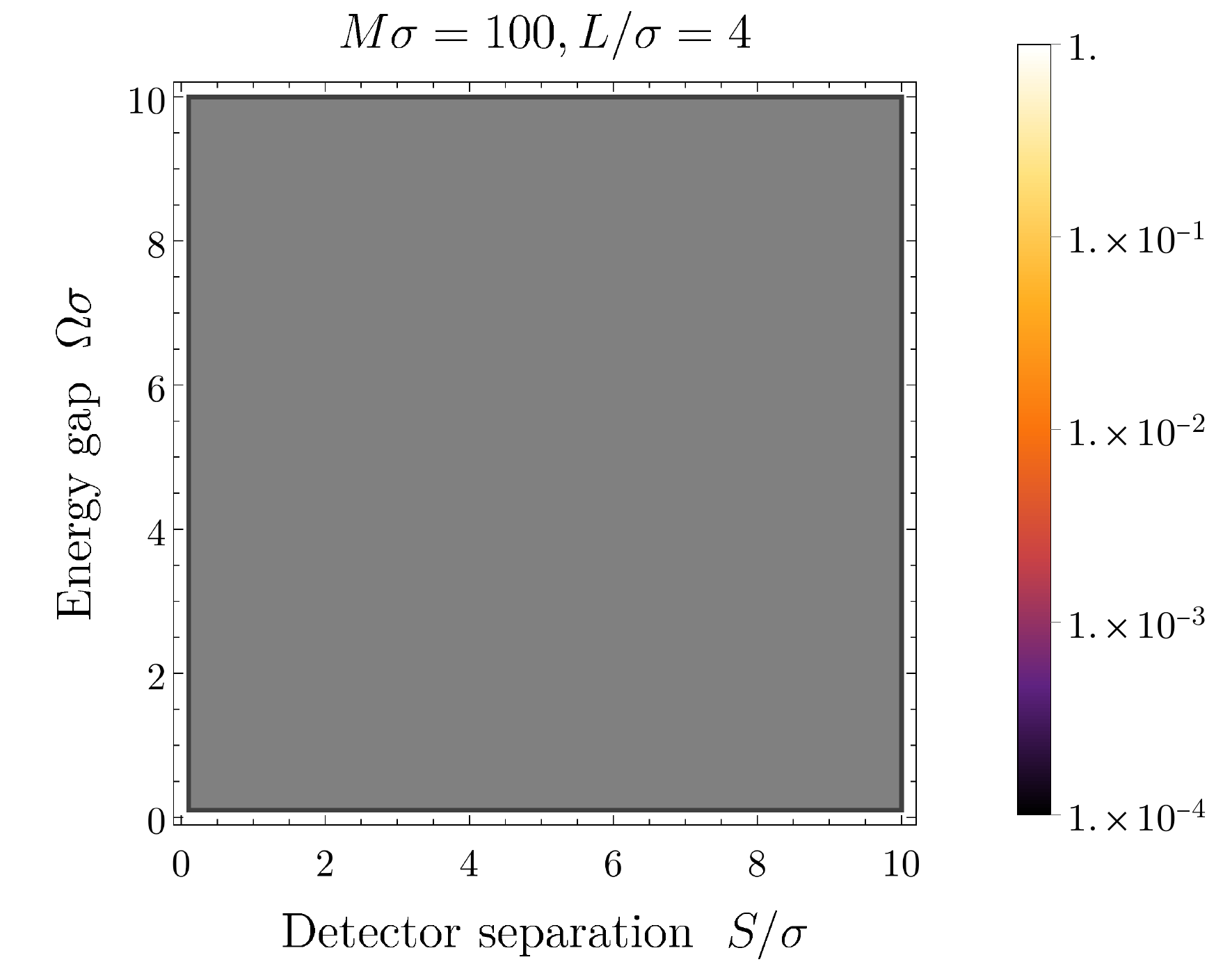}
    \hspace{0.8cm}
    \includegraphics[width=0.44\textwidth]{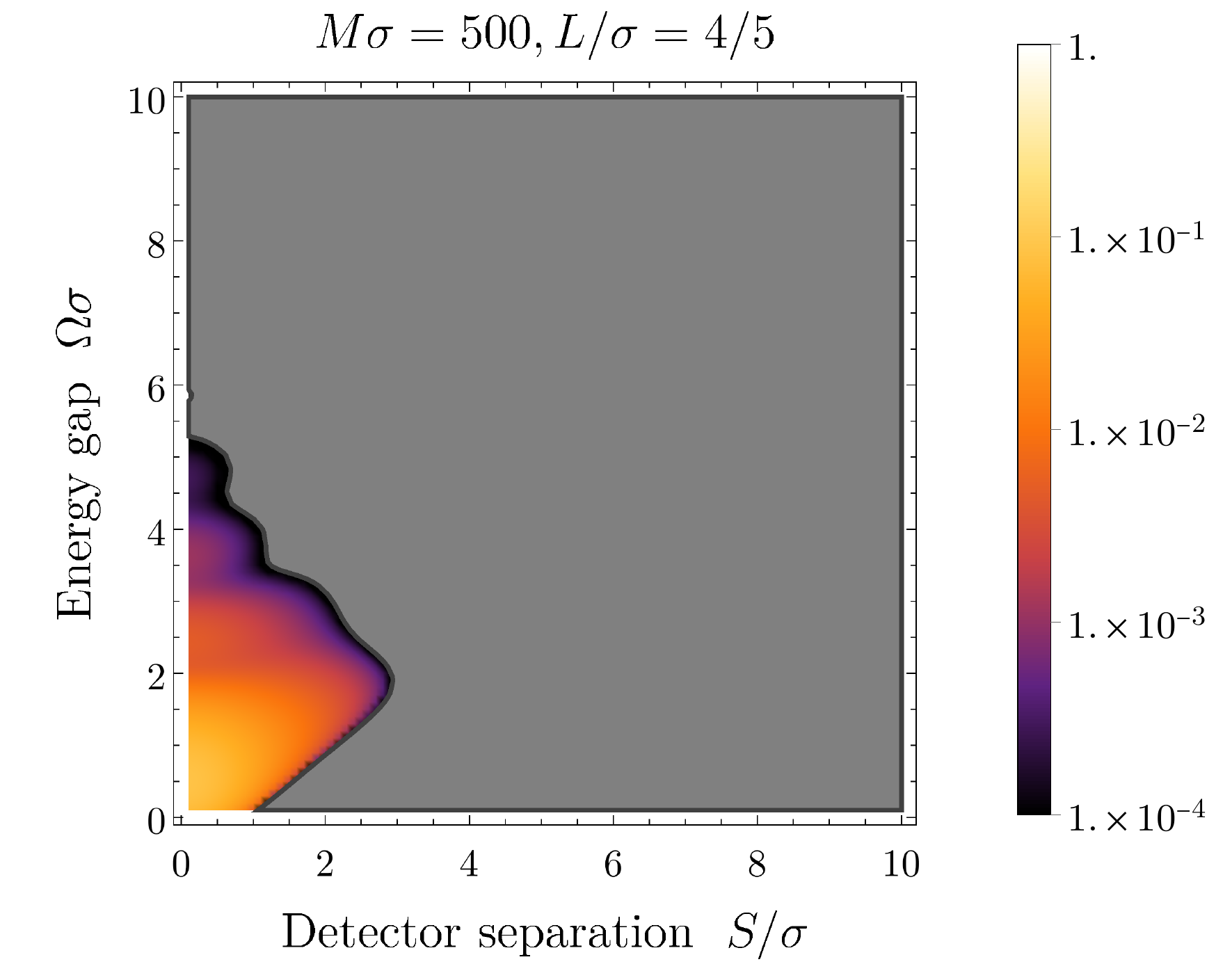}
    \par \vspace{0.8cm}
    \includegraphics[width=0.44\textwidth]{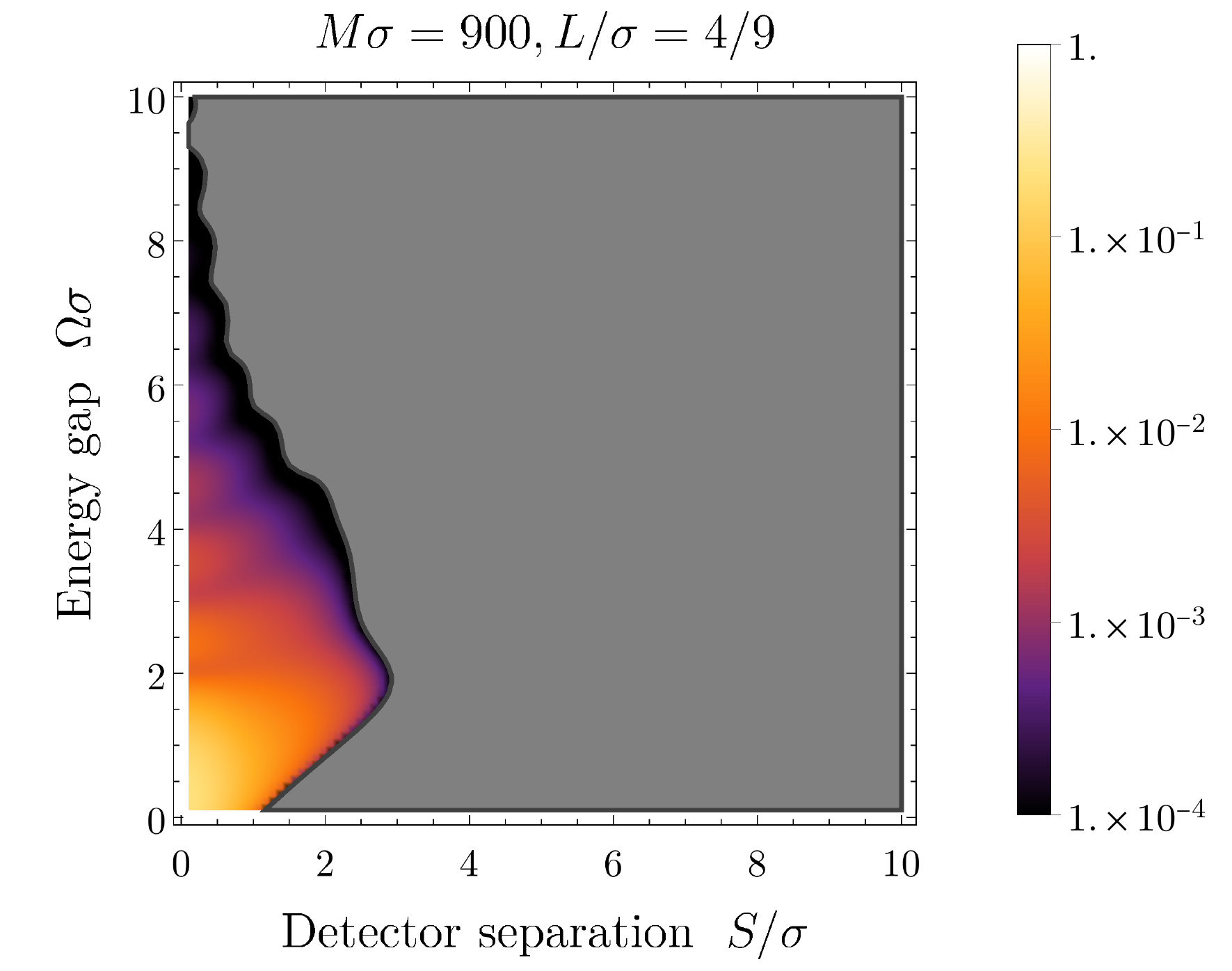}
    \hspace{0.8cm}
    \includegraphics[width=0.44\textwidth]{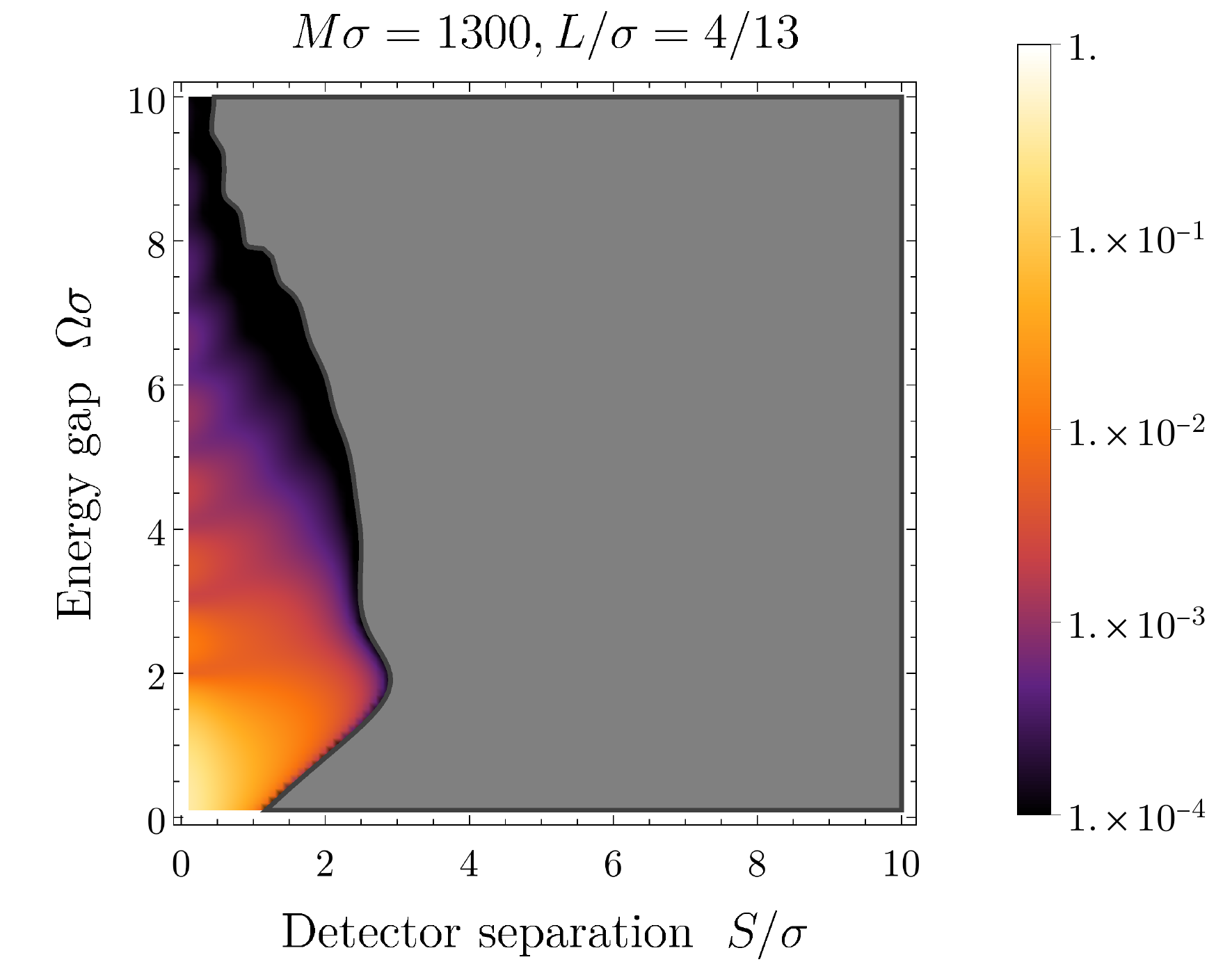}
    \par \vspace{0.8cm}
    \includegraphics[width=0.44\textwidth]{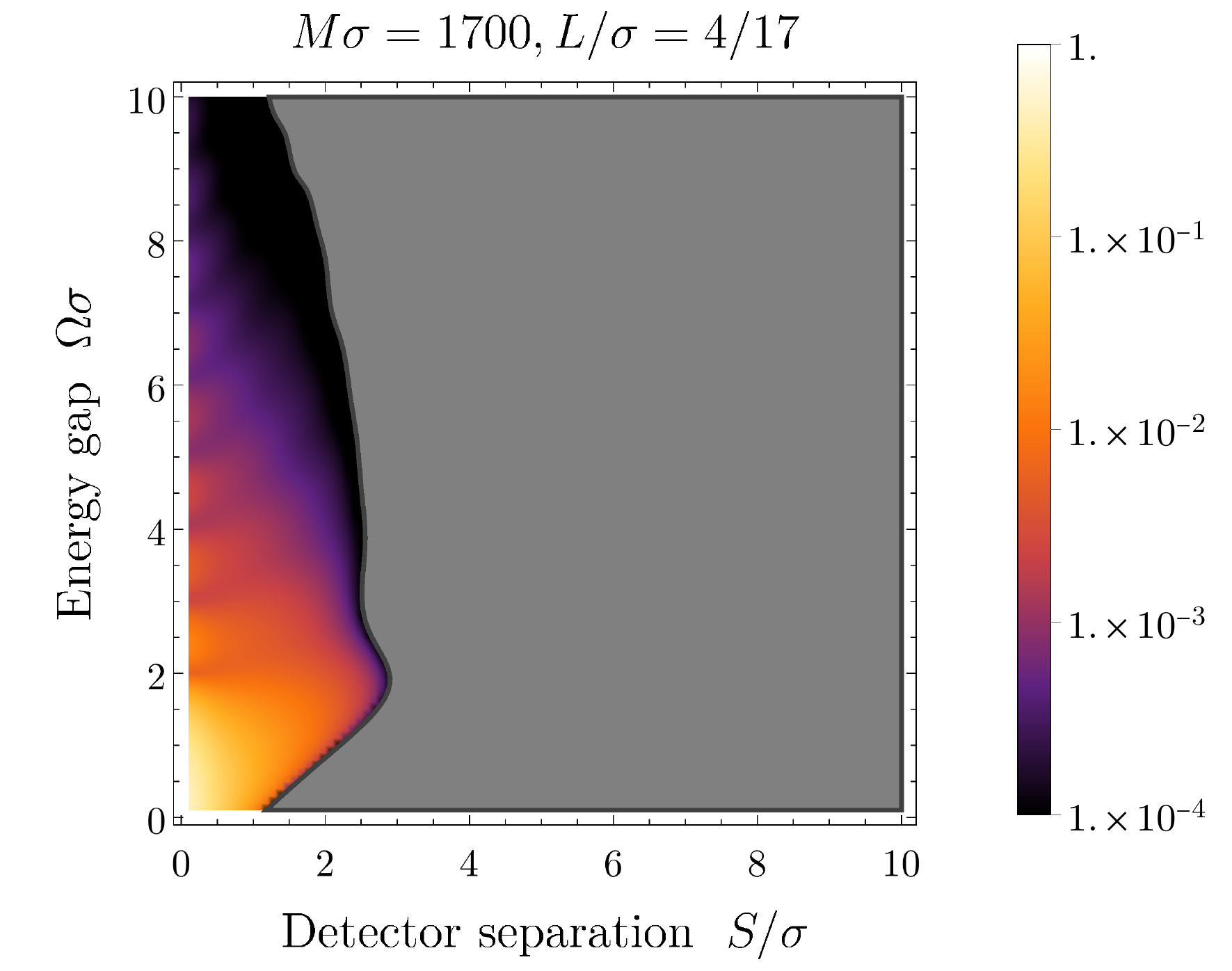}
    \hspace{0.8cm}
    \includegraphics[width=0.44\textwidth]{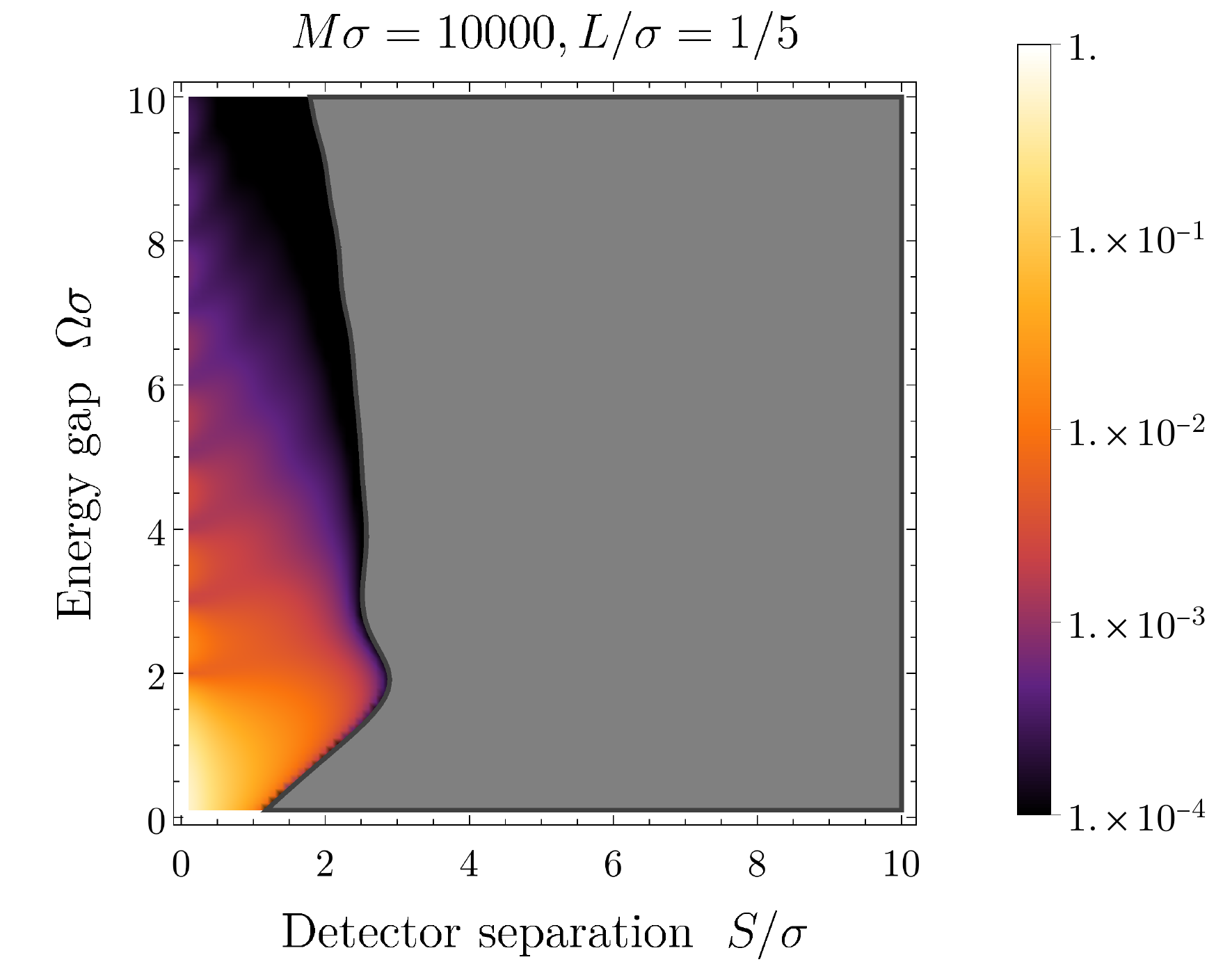}
    \par \vspace{0.3cm}
    \caption{
        The negativity $\mathcal{N}$ for two coherently delocalizing detectors, plotted as a function of the energy gap $\Omega$ and the separation $S$ of the two detectors.
        Regions of zero negativity are marked in grey.
        We chose the detector masses $M$ and the initial center of mass localization widths $L$ so that $\gamma$ decreases from left to right and from top to bottom.
        In the first five plots we fixed $1/(lmc)=2.5\times 10^{-3}$, while in the sixth plot we chose parameters satisfying $1/(lmc)=5\times 10^{-4}$, such as to see what happens to the negativity as we further decrease $1/(lmc)$. As expected, as we approach the limit $\gamma\to 0$ and $1/(lmc)\to 0$, we find that the negativity resembles more and more the negativity displayed in Fig.(\ref{fig:Negativity_UdW_detectors}) for two pointlike UdW detectors.}
    \label{fig:Negativity_massive_detectors}  
\end{figure*}

Let us now recall that Gaussians of width $L$ approach delta distributions in the limit $L\to 0$. Clearly, we should be able to recover the entanglement harvesting results for a pair of pointlike UdW detectors, in the limit of very large detector masses and center of mass distributions which are very sharply peaked (and thus essentially completely localized) at all times. However, we need to approach this limit in a way that ensures that the virtual center of mass velocities stay within the non-relativistic regime identified in Eq.(\ref{eq:non-relativistic-regime}). To this end, we define $M=:m/\gamma$ and $L=:l\gamma$, with $\gamma$ a regularization factor and with $m$ and $l$ constants satisfying $lmc \gtrsim 3.5 \times 10^2$. Letting $\gamma\to 0$ then lets the initial center of mass localization become very sharp ($L\to 0$) and the detector masses become very large ($M\to\infty$), while keeping the virtual center of mass velocities fixed and therefore non-relativistic. In the limit $\gamma\to 0$, the excitation probabilities and the entangling term reduce to
\ba
    &P& 
\to
    P^{\,c}_{0} 
    + \frac{\lambda^2 \sigma^2}{4 \pi^2 c (lmc)^2 }
    \int_0^{\infty} dk \, \frac{\sigma^2 c^2 k^3}{((\Omega + ck)^2 \sigma^2 -1)^4}
    \nonumber
    \\
&&\times
    \Big(
    20 (\Omega + ck)^2 \sigma^2 + 4
    + \cos(\pi\sigma (\Omega + ck))
    \nonumber \\
&&\times
    \left[
    20 (\Omega + ck)^2\sigma^2 
    + 4 - \pi^2 ((\Omega + ck)^2 \sigma^2-1)^2 
    \right]
    \nonumber
    \\
&&
    + 8(\Omega + ck)\pi\sigma[(\Omega + ck)^2 \sigma^2-1 ]\sin(\pi\sigma (\Omega + ck))
    \Big)
    \nonumber
    \\
&&
    + \mathcal{O}\left((lmc)^{-4} \right)
    \,,
    \\
&\mathcal{M}&
    \to
    \mathcal{M}^{\,c}_{0}
    + \frac{\lambda^2 \sigma^2 }{4 \pi^2 c (lmc)^2}
    \int_0^{\infty} dk  
    \frac{\sin(kS)}{S} D(k)
    \nonumber
    \\
    && + \mathcal{O}\left((lmc)^{-4} \right)
    \,.
\ea
By letting the virtual center of mass velocities go to zero, we can then describe two detectors whose center of mass degrees of freedom are localized very sharply at all times. Indeed, taking the limit $\gamma\to 0$ first and then taking the limit $1/(lmc)\to 0$, we recover the excitation probabilities and the entangling term for two pointlike UdW detectors, $P \to P^{\,c}_{0}$ and $\mathcal{M} \to \mathcal{M}^{\,c}_{0}$. 
We hence identified the limit in which entanglement harvesting for a pair of coherently delocalized detectors reduces to entanglement harvesting for a pair of pointlike UdW detectors. On the other hand, we find that there is no limit in which the results reduce to entanglement harvesting for a pair of spatially smeared UdW detectors. This confirms what we mentioned before, namely that spatial smearing profiles are appropriate to model the finite spatial extent of atoms due to their electronic orbitals \cite{Edu_Alejandro_harvesting2}, but not to model the coherent center of mass delocalization of an atom \cite{Stritzelberger_Kempf}.

In Fig.(\ref{fig:Negativity_massive_detectors}), we plotted the entanglement negativity for two coherently delocalizing detectors, as a function of the energy gap and the separation of the detectors. We can clearly see how the negativity reduces to the negativity for a pair of pointlike UdW detectors, when first letting $\gamma\to 0$ and then also letting $1/(lmc)\to 0$. We also observe that entanglement harvesting is indeed highly suppressed in the initial center of mass delocalization width.

Overall, we find that entanglement harvesting is suppressed for coherently delocalized detectors (and thus for actual physical matter systems such as atoms, ions or molecules), compared to entanglement harvesting for UdW detectors, whose center of mass degrees of freedom are assumed to be classical. 
An intuitive explanation for this suppression might be the following. 
We here focused on the entanglement harvested by the internal degrees of freedom of the two detectors. However, further (bipartite as well as multipartite) entanglement could potentially build up between the respective internal and center of mass degrees of freedom of the two detectors. 
This entanglement, which remains to be calculated, might build up at the expense of entanglement between the internal degrees of freedom of the two coherently delocalized detectors.

\section{Entanglement harvested by coherently delocalized detectors from the ground state of a medium}

Experimentally verifying entanglement harvesting from the vacuum is a difficult task \cite{Ralph_vacuum_harvesting1, Ralph_vacuum_harvesting2,Vacuum_Harvesting_experiment1, Vacuum_Harvesting_experiment2}. It might be more feasible to experimentally observe entanglement harvesting from the ground state of a medium, e.g., by sending atoms through a thin foil or a Bose-Einstein condensate. We here want to shed some light on whether the internal degrees of freedom of quantum delocalized atoms might become entangled with each other, via their respective interaction with the entangled ground state of a medium. 

\begin{figure}
    \includegraphics[width=0.46\textwidth]{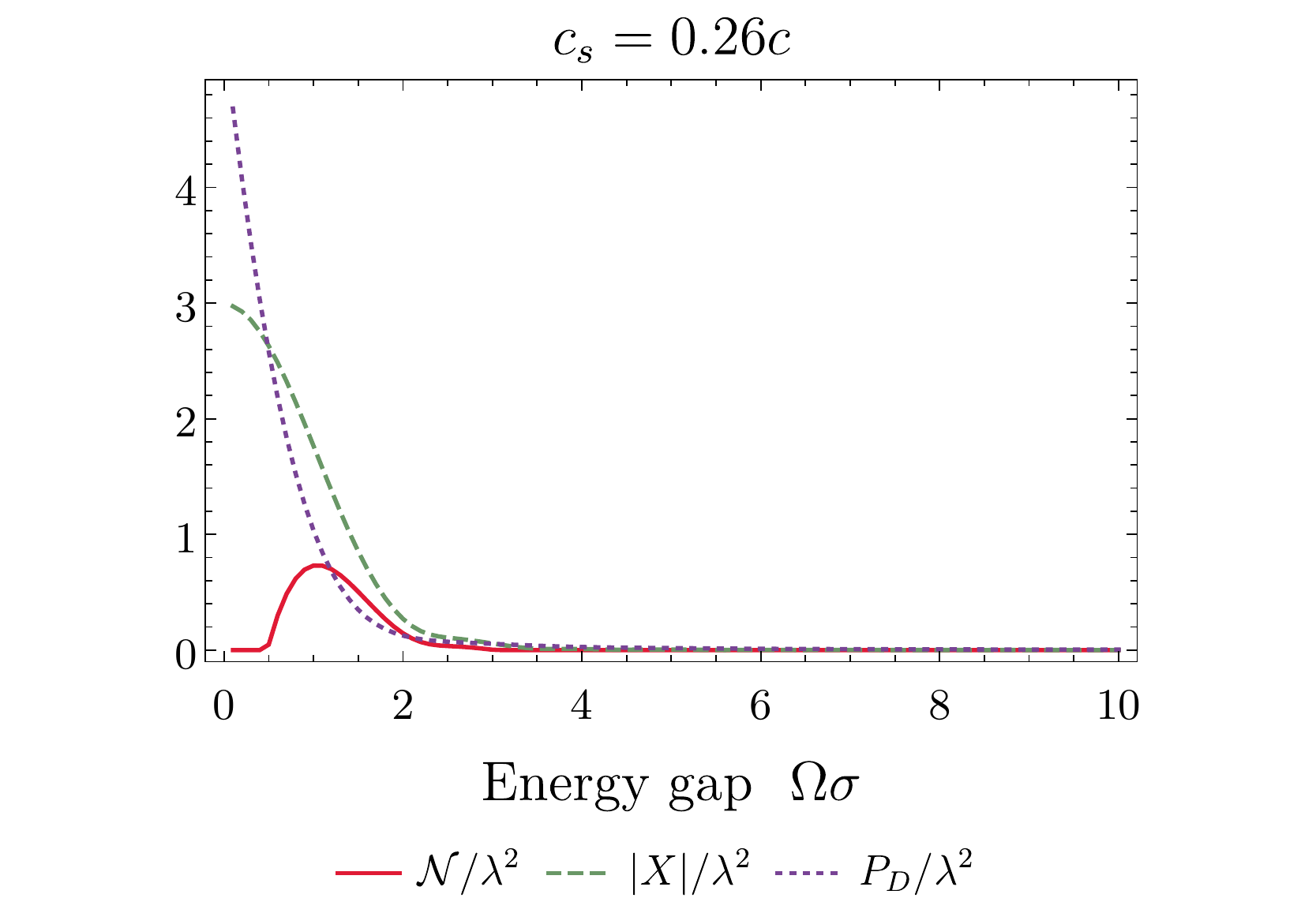}
    \par \vspace{0.8cm}
    \includegraphics[width=0.45\textwidth]{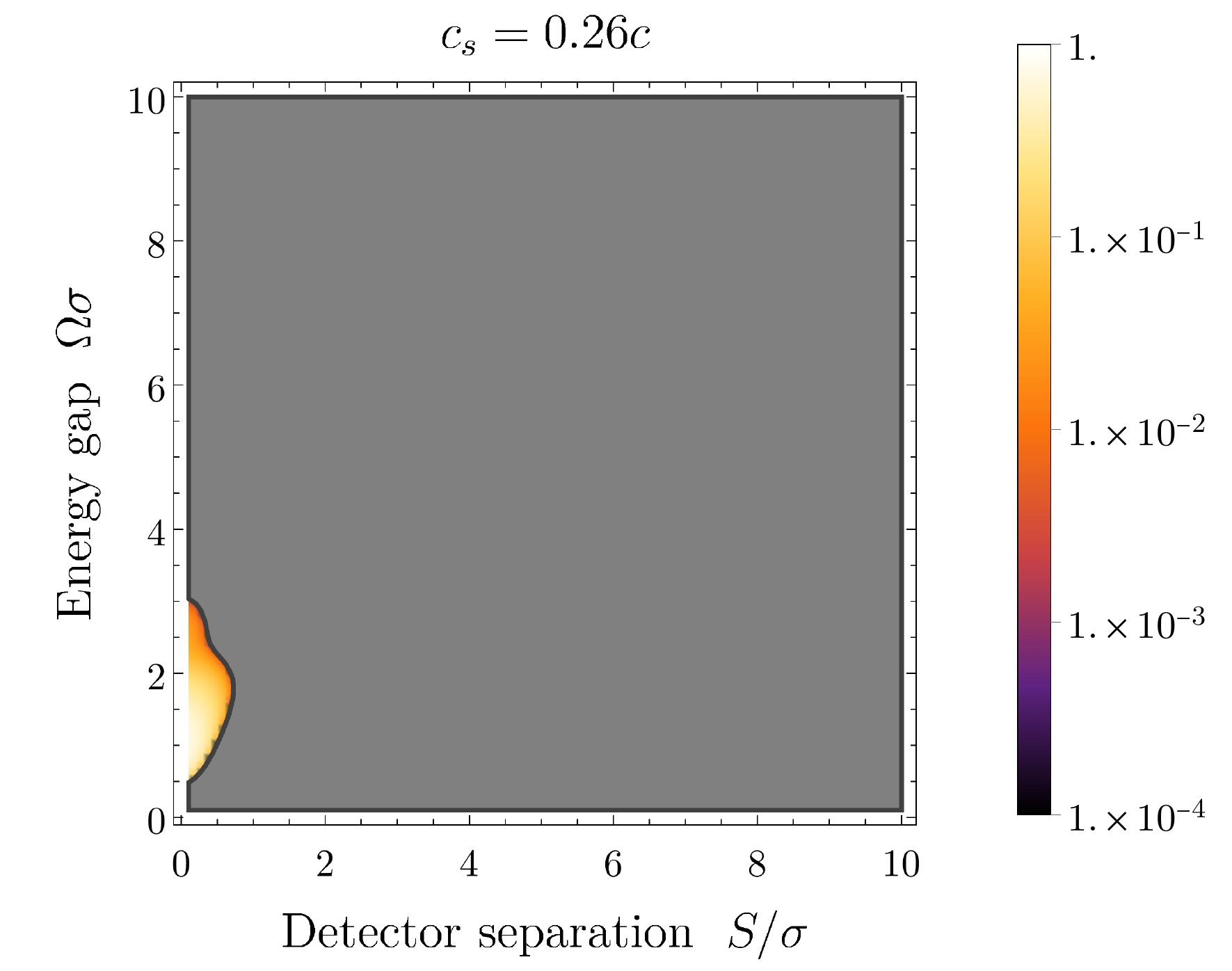}
    \caption{
        We consider two detectors (with detector masses $M\sigma=900$ and initial localization widths $L\sigma=4/9$) in a medium with wave propagation speed $c_s=0.26c$. We plot (\emph{top}) the transition probability $P$, the entangling term $\mathcal{M}$ and the negativity $\mathcal{N}$, as function of the energy gap $\Omega$ and for a detector separation $S=\sigma/10$, and (\emph{bottom}) the negativity $\mathcal{N}$ as a function of the energy gap and the detector separation. The region of zero negativity is marked in grey.}
    \label{fig:Negativity_medium2}  
\end{figure}

In the previous sections, we modeled the electromagnetic field via a simple scalar quantum field with dispersion relation $\omega = ck$, where $c$ stands for the vacuum propagation speed of light.
We will here model a medium via a scalar quantum field with dispersion relation $\omega = c_s k$, with $c_s < c$ the wave propagation speed in the medium. The propagation of waves in the scalar field could then for instance model the propagation of light in a medium, or the propagation of sound in a phononic field, both of which are known to propagate slower than light in the vacuum.
For concrete experimental setups, it will be very interesting to pursue analogous calculations with the there relevant realistic dispersion relation. 
Repeating the calculations we performed in the previous section, we obtain the excitation probabilities 
\ba
    P
&=&
    \frac{\lambda^2 \sigma^2}{4 \pi^2 c_s} 
    \bigg[ \mathcal{U}(0) + \frac{3}{2L^2} \frac{\partial^2 \mathcal{U}}{\partial p^2} (0)
    + \mathcal{O}\left( (LMc)^{-4} \right) \bigg] 
    \qquad 
\ea
and the entangling term
\ba
    \mathcal{M}
&=&
    \frac{\lambda^2 \sigma^2}{4\pi^2 c_s} 
    \int_0^{\infty} dk  
    \frac{\sin(kS)}{S} 
    e^{-L^2 k^2/4}
    \bigg[
    \mathcal{V}(k,0,0)
    \nonumber
    \\
    && 
    + \frac{1}{2L^2}\left[ \frac{\partial^2 \mathcal{V} }{\partial p_1^2} (k,0,0)
    + \frac{\partial^2 \mathcal{V} }{\partial p_2^2}(k,0,0) \right]
    \nonumber
    \\
    &&
    + \mathcal{O}\left( (LMc)^{-4} \right)
    \bigg]
    \,,
\ea
where $\mathcal{U}$ and $\mathcal{V}$ are defined as in Eq.(\ref{eq:U}) and Eq.(\ref{eq:V}), with the exception that $c$ is being replaced by $c_s$ in the definitions of $\mathcal{U}$ and $\beta_j$.

\begin{figure}
    \centering
    \includegraphics[width=0.46\textwidth]{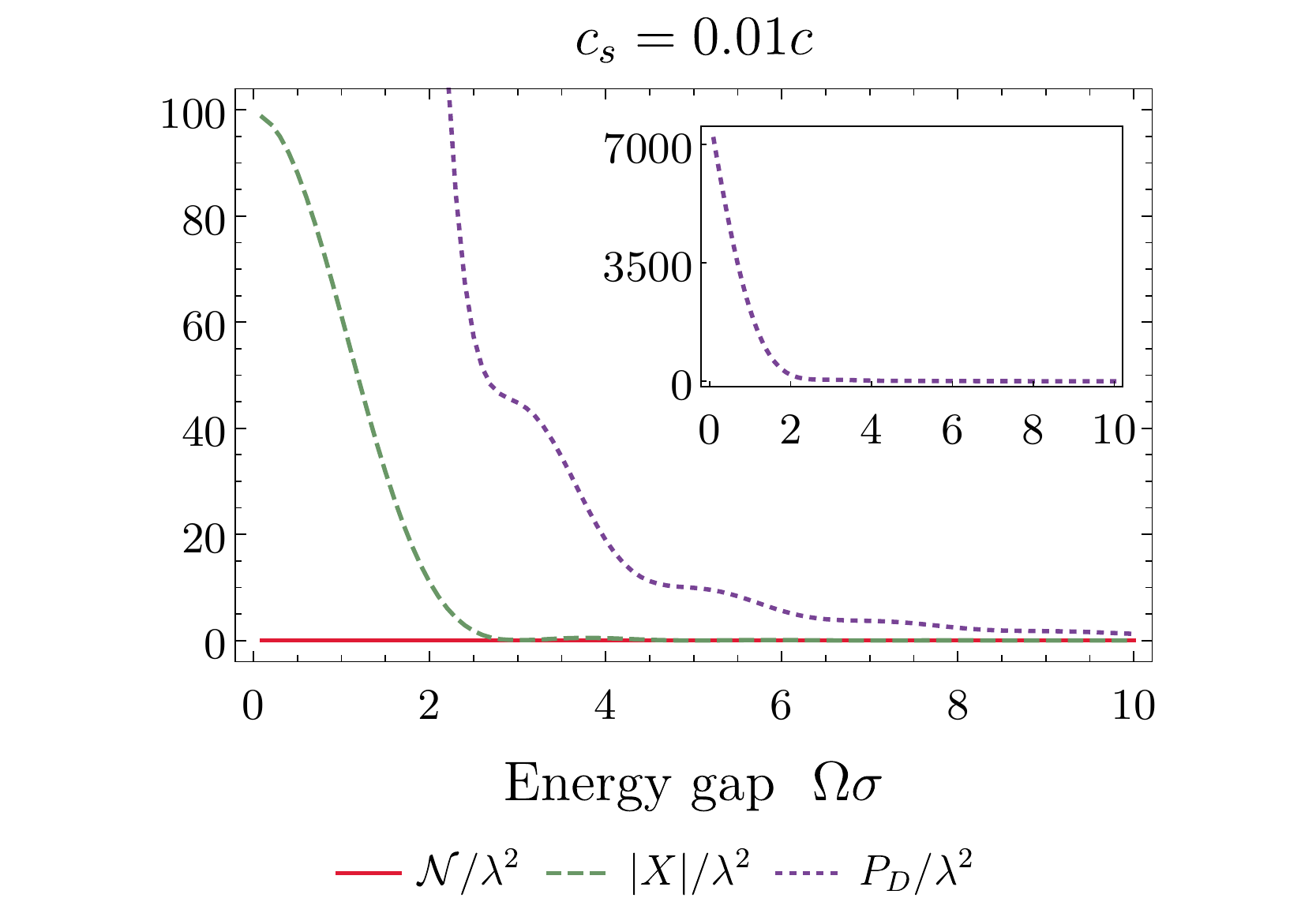}
    \caption{
        We consider two detectors (with detector masses $M\sigma=900$, initial localization widths $L\sigma=4/9$ and detector separation $S=\sigma/10$) in a medium with wave propagation speed $c_s=0.01c$. We plot the transition probability $P$ and the entangling term $\mathcal{M}$ as function of the energy gap, and we find that the negativity $\mathcal{N}$ vanishes for this choice of parameters.
        }
    \label{fig:Negativity_medium}  
\end{figure}

In Fig.(\ref{fig:Negativity_medium2}) we plotted the excitation probability, the entangling term and the negativity for a pair of coherently delocalized detectors in a medium with wave propagation speed $c_s=0.26c$. We find that the negativity is significantly suppressed, compared to the negativity in Fig.(\ref{fig:Negativity_massive_detectors}), in which the detectors were in the vacuum. 
Intuitively, this behavior can be explained as follows: transforming into the quantum uncertain rest frame of the delocalizing detectors, the phononic ground state transforms non-trivially into an excited field state which might be more entangled than the phononic ground state. 
The entangling excitations thus potentially increase, but at the same time, also the local ``noisy" excitations increase. For the parameters we chose here, these two competing effects play out in such a way that the negativity decreases significantly. 
Both light and sound can be slowed down significantly in media (e.g. light in crystals or sound in Bose-Einstein condensates), to the extreme of being stopped completely \cite{stopped_light, BEC_speed_of_sound}. 
The detectors in such media could coherently delocalize with virtual velocities that are comparable to, or even larger than, the propagation speed in the medium, $v\gtrsim c_s$, while remaining well within the non-relativistic regime, $v\leq 0.01c$. Gaussian center of mass wave packets with support for supersonic virtual center of mass velocities are ones for which $LMc_s \gtrsim 3.5$, while the non-relativistic regime is characterized by $LMc\gtrsim 3.5 \times 10^2$.
In Fig.(\ref{fig:Negativity_medium}) we plotted the excitation probabilities, the entangling term and the negativity for two detectors in a medium whose wave propagation speed is $1\%$ of the vacuum speed of light. We chose the parameters so that the maximal virtual center of mass velocities in the Gaussian wave packet are close to the speed of sound in the medium  ($LMc_s=4$), while staying well within the non-relativistic regime ($LMc=400$). We find that both the entangling term and the excitation probabilities for detectors in the medium are significantly enhanced, and we find that overall the negativity vanishes.
We thus find that if the wave propagation speed in the medium is too small, the internal degrees of freedom of a pair of coherently delocalizing detectors cannot become entangled with each other.

We conjecture that it is generally harder for detectors to harvest entanglement from a medium than from the vacuum. Entanglement harvesting experiments in media might however still be worth considering, given that they may be more easily conducted than the harvesting of entanglement from the vacuum. 

\section{Conclusions and Outlook}

We here studied the ability of two quantum delocalized detectors to become entangled with one another, via their respective interaction with a scalar quantum field.
Specifically, we accounted for the quantum mechanical nature of the center of mass degrees of freedom of the detectors, and calculated the entanglement negativity for the internal degrees of freedom of the two  detectors. Our calculations show that the process of entanglement harvesting is affected by the coherent delocalization of matter and, in particular, that delocalized detectors harvest less entanglement than detectors whose center of mass degrees of freedom are assumed to behave classically.

We identified the limit in which the results for entanglement harvesting for coherently delocalized detectors reduce to the results for detectors with classical external degrees of freedom: For two detectors of very large mass, whose center of mass wave functions are initially very sharply peaked and which dynamically delocalize very slowly, we recover the negativity for two pointlike UdW detectors.
This limit corresponds to detectors whose centers of mass are essentially completely localized at all times. 

Further, we found that the delocalization of the centre of mass degree of freedom is fundamentally different from the finite extent of a detector's charge distribution that arises from the finite size of its electronic orbitals. If the description of the latter is desired, it can be modeled separately through the use of smearing functions. Finally, we discussed entanglement harvesting in media, where we found that entanglement harvesting for coherently delocalized detectors decreases with decreasing wave propagation speeds.

Looking forward, it could be useful to adapt and apply methods of quantum reference frames \cite{QRF_Aharonov_Susskind,QRF_Aharonov, QRF_Bartlett_Rudolph_Spekkens, QRF_Giacomini_Castro-Ruiz_Brukner, QRF_Palmer_Girelli_Bartlett} to the study of  entanglement harvesting. On one hand, these methods may allow one to extend our study to relativistic virtual center of mass velocities. On the other hand
it might also be insightful to apply the quantum reference frame formalism to entanglement harvesting for coherently delocalized detectors in media. One could for instance consider the ground state of a medium and perform a quantum reference frame transformation into the quantum uncertain rest frame of the detectors. 
One might then explicitly see whether the ground state transforms into an excited field state, and whether this excited field state is more entangled than the ground state. 
One could then try to apply these insights towards finding  quantum uncertain detector states of motion that are best suited for the purpose of optimally harvesting entanglement from the medium. 

A deeper understanding of entanglement harvesting in the vacuum and in media may also be obtainable through the study of entanglement harvesting from excited field states. Intuitively, states with field quanta delocalized over large distances may contain harvestable entanglement over large distances. The  intuition here is nontrivial, however, because while the entangling term might increase, the excitation probabilities might also increase significantly. Overall, their difference, the entanglement negativity, might therefore decrease, similarly to what we already observed in this paper for detectors in media. 

Further, in this paper we focused on the entanglement harvested only by the internal degrees of freedom of quantum delocalized detectors. It will be very interesting to investigate to what extent the center of mass degrees of freedom of coherently delocalized detectors can harvest entanglement from the vacuum. In addition, the center of mass degrees of freedom can become entangled with the internal degrees of freedom in the harvesting process. We conjecture, for example, that for faster virtual recoil velocities, the center of mass degree of freedom harvest larger amounts of entanglement while, possibly due to entanglement monogamy, the internal degrees of freedom then might harvest less.

It will be technically difficult, however, to calculate entanglement measures, such as the negativity, for the center of mass degrees of freedom, since they possess Hilbert spaces that are infinite dimensional. We anticipate that this can be addressed, for example, by either postselecting for specific recoil momenta, by discretizing the momentum space, e.g., by placing the detectors in a confining potential or cavity and placing an energy cutoff, or by binning momenta into a finite number of momentum regions. 

Such methods could then allow one, for example, to study to what extent, e.g., a pre-existing entanglement between the center of mass degrees of freedom can help or hinder the harvesting of entanglement. It should also be very interesting to explore how the delocalization of the center of mass interacts with models of natural ultraviolet cutoffs in quantum field theory, such as bandlimitation, see, e.g., \cite{Pye_2015, henderson2020bandlimited, achim_bandlimit}. 

We close by highlighting that, after the harvesting process, the extracted entanglement is available for arbitrary purposes, including use in protocols that also involve two detectors interacting with a quantum field, such as quantum state teleportation and quantum energy teleportation. It should be interesting to explore if these protocols could be usefully merged with entanglement harvesting into one protocol for, e.g., quantum state or energy teleportation without the need of pre-existing entanglement. 

\vspace{\baselineskip} 
\vspace{\baselineskip}

\begin{acknowledgments}

The authors acknowledge useful discussions with Tim Ralph, Eduardo Mart\'in-Mart\'inez, Robert~B.\ Mann, Giacomo Pantaleoni, and Ben Baragiola. 
AK and NS are grateful to Tim Ralph for his kind hospitality at the University of Queensland.
N.S.\ acknowledges support through an Ontario Trillium scholarship and a Mitacs Globalink Research Award. 
L.J.H.\ acknowledges support from the Natural Sciences and Engineering Research Council of Canada (NSERC). 
A.K.\ acknowledges support through the Discovery Program of the National Science and Engineering Research Council of Canada (NSERC) and through two Google Faculty Awards.
This work is supported by the U.S. Air Force Research Laboratory Asian Office of Aerospace Research and Development (Grant No.\ FA2386-16-1-4020), the Australian Research Council~(ARC) Discovery Program (Project No.\ DP200102152), and the ARC Centre of Excellence for Quantum Computation and Communication Technology (Project No.\ CE170100012).

\end{acknowledgments}

\bibliographystyle{ieeetr}
\bibliography{references}

\end{document}